\documentclass[
 aps,
 prd,
 reprint,
 superscriptaddress,
 nofootinbib,
 longbibliography
]{revtex4-2}

\usepackage{amsmath,amssymb}
\usepackage{booktabs}
\usepackage{graphicx}
\usepackage{mathrsfs}
\usepackage{capt-of}
\usepackage[hidelinks]{hyperref}
\hypersetup{
 pdftitle={Flat-limit of artificial cosmology for Schwarzschild--de Sitter scalar waves},
 pdfauthor={Anil Zenginoglu and Govind Arun Kumar},
 pdfsubject={Curvature-coupled scalar-wave evolution on Schwarzschild--de Sitter and Schwarzschild spacetimes},
 pdfkeywords={general relativity, black holes, numerical relativity, null infinity, wave propagation, conformal coupling}
}

\newcommand{\scri}{\mathscr{I}}
\newcommand{\hc}{\mathcal{H}_{c}^{+}}
\newcommand{\hb}{\mathcal{H}_{b}^{+}}

\begin{document}

\title{Flat limit of artificial cosmology for scalar waves in Schwarzschild--de Sitter}

\author{An\i l Zengino\u{g}lu}
\email[Corresponding author: ]{anil@umd.edu}
\affiliation{Institute for Physical Science and Technology, University of
Maryland, College Park, Maryland 20742, USA}

\author{Govind Arun Kumar}
\email{govind02@umd.edu}
\affiliation{Institute for Physical Science and Technology, University of
Maryland, College Park, Maryland 20742, USA}

\begin{abstract}
How do cosmological waveforms approximate waveforms from isolated systems for a small cosmological constant? This question is motivated by the observed small value of the cosmological constant, and also by Misner's idea to regulate the formally singular terms introduced into the Einstein equations by conformal compactification. We
investigate this question quantitatively using scalar waves on Schwarzschild--de Sitter backgrounds. A suitable flat limit requires the foliation to remain regular at the future conformal boundary. We compare bridge foliations extending from the black hole horizon to the cosmological horizon with a regular asymptotically flat limit. Comparisons of cosmological and asymptotically flat waveforms on geometrically matched data show that the flat limit is numerically well behaved and recovers the essential features of the asymptotically flat waveform. We extend Misner's artificial cosmology by introducing a cosmological source that leaves the near-zone geometry unchanged and show that it improves agreement of the ringdown waveform but fails to capture the intermediate polynomial tail behavior accurately.
\end{abstract}

\maketitle
\raggedbottom

\section{Introduction}
\label{sec:introduction}

Gravitational radiation from isolated systems is unambiguously defined at future null infinity, $\scri^+$. It is therefore important to include null infinity in the computational domain when calculating waveforms. In hyperboloidal compactification, we use spacelike slices that approach cuts of null infinity and solve for conformal variables rescaled by suitable powers of the conformal factor, $\Omega$. A difficulty in this approach is the treatment of formally singular terms at infinity. Misner proposed regulating such terms with a small positive cosmological constant \cite{misner_hyperboloidal_2006,misner_over_2006,misner_excising_2006}. At finite cosmological length $L$, we replace replace null infinity where $\Omega=0$ with a cosmological horizon where $\Omega_L>0$.  This raises to question how such a construction can approximate the asymptotically flat waveform. Another motivation for this work comes from the fact that cosmological observations are consistent with a small positive cosmological constant in our Universe \cite{Planck:2018cosmology}.  This gives a physical
motivation for numerically investigating how well asymptotically flat waveforms approximate those in a spacetime with small positive $\Lambda$.  We address this question using scalar waves in Schwarzschild--de Sitter (SdS) spacetimes.

There are various approaches to include the asymptotic boundary in a computational domain. Characteristic evolution and Cauchy--characteristic matching can now successfully calculate nonlinear binary-black-hole mergers and tails \cite{Winicour:2012characteristic,MaEtAl:2025CCM}.  The hyperboloidal approach uses spacelike slices instead \cite{friedrich1983cauchy}. Early calculations exposed difficulties related to constraints, gauge conditions, and moving boundaries \cite{Frauendiener:1998II,Hubner:1999scheme,Hubner:2001timelike,FrauendienerHein:2002,Husa:2002problems}.  More recently, generalized conformal field equations based on a Weyl connection and conformal geodesics have demonstrated global calculations and nonlinear black-hole scattering \cite{Friedrich:1995general,Friedrich:2003geodesics,
zenginouglu2007numerical, zenginouglu2007conformal, BeyerEtAl:2017GCFE,FrauendienerStevens:2021massloss,
FrauendienerStevensThwala:2025,CamdenEtAl:2025,
FrauendienerStevens:2023compactification}. While the coordinate location of the conformal boundary is a priori known in this approach, the problem of moving boundaries and the associated loss of resolution remains.

Scri fixing maps null infinity to a time-independent boundary by using a suitable conformal compactification on stationary hyperboloidal foliations \cite{zenginoglu_hyperboloidal_2008}, which proved very useful in black-hole perturbation theory. Time-domain Schwarzschild and Kerr calculations demonstrated accurate long evolutions, including power-law tails \cite{zengin2008hyperboloidal,ZenginogluNunezHusa:2009,
ZenginogluTiglio:2009,RaczToth:2011,Bernuzzi:2011aj,
ZenginogluKhanna:2011,HarmsBernuzziBrugmann:2013,PanossoMacedoAnsorg:2014,
Harms:2014dqa,CsukasRaczToth:2019}.
Applications also include particle-source evolutions
\cite{GomesDaSilvaEtAl:2023,VaswaniEtAl:2026}, second-order Kerr perturbations
\cite{RipleyEtAl:2021}, ringdown analysis \cite{ZhuEtAl:2024}, and free
3+1 evolutions \cite{ZenginogluKidder:2010, ZenginogluGalley:2012, yang2013quasinormal, Peterson:2023bha,Rinne:2025nonlinear,
ZenginogluBernuzziNutzi:2026}. This idea has also been applied in the
frequency domain to quasinormal-mode calculations
\cite{Zenginoglu:2011jz,AnsorgMacedo:2016,MacedoJaramilloAnsorg:2018,
Macedo:2020,PanossoMacedo:2024nkw,BessonJaramillo:2025}, pseudospectrum analyses
\cite{JaramilloMacedoAlSheikh:2021,GasperinJaramillo:2022,
DestounisBoyanovMacedo:2024}, and self-force calculations
\cite{MacedoEtAl:2022,MacedoEtAl:2024selfforce,Leather:2025selfforce,
LeatherEtAl:2026Teukolsky}.

Applying scri-fixing to nonlinear Einstein equations is a difficult problem due to formally singular terms \cite{Frauendiener:2000mk, Zenginoglu:2008pw}. Implementations have been demonstrated in spherical symmetry using conformal BSSN/Z4 systems
\cite{Vano-Vinuales:2014koa,Vano-Vinuales:2017qij,
VanoVinuales:2024reference,Alvares:2025emkg} and 
dual-foliation generalized harmonic gauge \cite{Hilditch:2015aba,HilditchEtAl:2018,DuarteEtAl:2023,
Peterson:2024spherical,Peterson:2025strong}. Beyond spherical symmetry, the only implementations of hyperboloidal scri-fixing use constrained evolution \cite{Moncrief:2008ie,Rinne:2009qx,RinneMoncrief:2013,
BuchmanPfeifferBardeen:2009,BardeenSarbachBuchman:2011}. Building a robust, free-evolution hyperboloidal code that
fixes $\scri^+$ for generic three-d`imensional spacetimes remains an open problem.

Misner's idea of using a small positive cosmological constant to regulate formally singular terms is an attractive alternative approach. This idea has only been tested in de Sitter spacetime with a Minkowski limit by Misner and collaborators \cite{misner_hyperboloidal_2006,misner_over_2006,misner_excising_2006}. The Minkowski case is special due to the Huygens' principle according to which the support of the solution is confined to the light cone. In this paper, we study Misner's proposal in the presence of a black hole. We investigate the flat limit of artificial cosmology in the context of SdS spacetimes using scalar waves as a toy problem. For $\Lambda>0$, the future conformal boundary is spacelike, with different asymptotic symmetries, radiative degrees of freedom, and fluxes from those at $\scri^+$ \cite{Ashtekar:2014zfa,Ashtekar:2015lla,Ashtekar:2015lxa}.  We ask: how does a cosmological calculation approximate an asymptotically flat solution?  We compare minimal and conformal scalar coupling, $\xi=0$ and $1/6$, separating common regulator behavior from dependence on the asymptotic field equation \cite{brady_radiative_1999}.

The flat limit changes the global causal and conformal
structure.  Writing $\Lambda=3/L^2$ and holding $M$ fixed, the SdS metric approaches Schwarzschild as $L\to\infty$. The asymptotic limit, however, is singular: the cosmological horizon moves to infinite areal radius, and the spacelike future conformal boundary becomes null \cite{Geroch:1969limits, Ashtekar:2014zfa,BugdenPaganini:2019,ZhouPanossoMacedo:2025}.
Comparing waveforms at the outer boundary therefore requires a controlled
limit of the foliation and the extraction surfaces.  A regular hyperboloidal
flat limit requires the limiting slices to meet distinct cuts of future
null infinity, with a smooth, nondegenerate conformal spatial metric
\cite{zenginoglu_bridge_2025,ZhouPanossoMacedo:2025}.

As a dynamical demonstration of this singular limit, consider late-time decay. Schwarzschild fields have inverse-power tails \cite{price1972nonspherical}, whereas SdS fields decay exponentially \cite{Brady:1996za,brady_radiative_1999,Molina:2003dc}. Recovering a Schwarzschild signal over a finite interval is therefore different from recovering its asymptotic decay.  We test how far finite-time agreement extends as $L$ increases.
Building on the bridge construction of \cite{zenginoglu_bridge_2025} and the SdS limiting geometry of
\cite{ZhouPanossoMacedo:2025}, we give a quantitative time-domain test of artificial cosmology. To potentially improve Misner's proposal, we also test a varying cosmological source supported in the asymptotic domain based on a construction by Dymnikova \cite{Dymnikova:2000algebraic,Dymnikova:2002mass}. 

Section~\ref{sec:bridges} reviews the bridge foliations of SdS and their flat limit, Sec.~\ref{sec:scalar} defines the scalar comparison, and Secs.~\ref{sec:results} and~\ref{sec:exterior-supported} give the uniform and exterior-supported results.  The appendices present numerical details.

\section{Flat limit of SdS bridges}
\label{sec:bridges}

\subsection{SdS static patch}

The SdS metric in static coordinates is
\begin{align}
 ds^2&=-f(r)dt^2+\frac{dr^2}{f(r)}+r^2d\omega^2,
 \label{eq:sds_metric}\\
 f(r)&=1-\frac{2M}{r}-\frac{r^2}{L^2},
 \qquad d\omega^2=d\theta^2+\sin^2\theta\,d\varphi^2.
 \label{eq:f}
\end{align}
We use geometrized units $G_{\rm N}=c=1$. The coordinates are given in units of $M$. For $0<M<L/(3\sqrt{3})$, $f$ has a black-hole root $r_b$, a cosmological root $r_c$, and a negative root $r_u=-(r_b+r_c)$.  They satisfy
\begin{align}
 L^2&=r_b^2+r_br_c+r_c^2, \qquad
 2ML^2=r_br_c(r_b+r_c),
 \label{eq:root_relations}\\
 f(r)&=\frac{(r-r_b)(r_c-r)(r-r_u)}{L^2r}.
 \label{eq:f_factorized}
\end{align}
For $i=b,c$, the surface gravities follow from
\eqref{eq:kappa}.  We use the positive algebraic coefficient $\kappa_u:=|f'(r_u)|/2$ at the negative root:
\begin{equation}
 \kappa_i=\frac12\left|f'(r_i)\right|
 =\frac12\left|\frac{1}{r_i}-\frac{3r_i}{L^2}\right|,
 \qquad i\in\{b,c,u\}.
 \label{eq:kappa}
\end{equation}
Using a reference radius $r_0\in(r_b,r_c)$, the normalized
tortoise coordinate is
\begin{align}
r_{*,L}(r)={}&\frac{1}{2\kappa_b}
 \ln\frac{r-r_b}{r_0-r_b}
-\frac{1}{2\kappa_c}
 \ln\frac{r_c-r}{r_c-r_0}
\notag\\
&+\frac{1}{2\kappa_u}
 \ln\frac{r-r_u}{r_0-r_u}.
\label{eq:tortoise}
\end{align}
The tortoise coordinate goes to $-\infty$ at $r_b$ and to $+\infty$ at
$r_c$.  In the large-$L$ limit, the static patch expands to the Schwarzschild exterior \cite{ZhouPanossoMacedo:2025} with
\begin{equation}
 r_b\to2M,\qquad r_c=L-M+\mathcal{O}(M^2/L).
 \label{eq:horizon_flat_limit}
\end{equation}

\subsection{Global bridge foliation}

For a regular stationary foliation, we introduce a
stationary height function \cite{zenginoglu_hyperboloidal_2008,zenginoglu_bridge_2025},
\begin{equation}
 \tau=t+h(r),\qquad B(r):=f(r)h'(r),
 \label{eq:height}
\end{equation}
where $B$ is the boost relative to the tortoise coordinate $r_*$.  For a radial coordinate
$\rho=g(r)$ and $G=d\rho/dr$, the metric becomes
\begin{equation}
 ds^2=-f\,d\tau^2+\frac{2B}{G}\,d\tau d\rho
 +\frac{1-B^2}{fG^2}\,d\rho^2+r^2d\omega^2.
 \label{eq:transformed_metric}
\end{equation}
A future-directed spacelike bridge within the static patch satisfies
\begin{equation}
 |B|<1,\qquad B(r_b)=+1,\qquad B(r_c)=-1,
\label{eq:boost_conditions}
\end{equation}
At each horizon, the radial metric coefficient $(1-B^2)/(fG^2)$ must
approach a finite positive value.  For each finite $L$, our radial map has
nonzero boundary values of $G$, so this is equivalent to requiring a finite
positive limit of $(1-B^2)/f$.  
The horizons give $h\sim+r_*$ at $\hb$ and
$h\sim-r_*$ at $\hc$ \cite{zenginoglu_bridge_2025}. Note that these are local conditions at the horizons, and do not determine the foliation beyond $\hc$.  On the future
branch of the expanding region, where $f<0$, spacelike continuation beyond the cosmological horizon requires
$B<-1$.  How $B$ continues from its horizon value $B(r_c)=-1$ determines how the leaves approach the spacelike future boundary.  

Two height functions demonstrate this remaining freedom.  Both are normalized
by $h(r_0)=0$ and have the same horizon data:
\begin{align}
h_{\rm min}(r)={}&
\frac{1}{2\kappa_b}\ln\left|\frac{r-r_b}{r_0-r_b}\right|
+\frac{1}{2\kappa_c}\ln\left|\frac{r_c-r}{r_c-r_0}\right|,
\label{eq:minimum_height}\\
h_{\rm mg}(r)={}&
\frac{1}{2\kappa_b}
\ln\left|\frac{(r-r_b)r_0}{(r_0-r_b)r}\right|
+\frac{1}{2\kappa_c}
\ln\left|\frac{(r_c-r)r_0}{(r_c-r_0)r}\right|
\notag\\
&-\frac{1}{2\kappa_u}
\ln\left|\frac{(r-r_u)r_0}{(r_0-r_u)r}\right|.
\label{eq:minimal_height}
\end{align}
The two logarithms in the minimum height function $h_{\rm min}$ are sufficient for horizon regularity. The second height function comes from the minimal gauge which was first discussed in hyperboloidal QNM calculations \cite{AnsorgMacedo:2016,MacedoJaramilloAnsorg:2018, Macedo:2020}.  Its derivation for static spherical metrics uses the in-out and out-in strategies of \cite{PanossoMacedo:2023qzp}.  For SdS, the out--in construction adds the regular third-root contribution and gives a regular Schwarzschild limit \cite{ZhouPanossoMacedo:2025}.

\begin{figure*}[t]
 \centering
 \begin{minipage}[t]{0.49\textwidth}
  \centering
  \textbf{(a) Minimum height}\\
  \includegraphics[width=\linewidth]{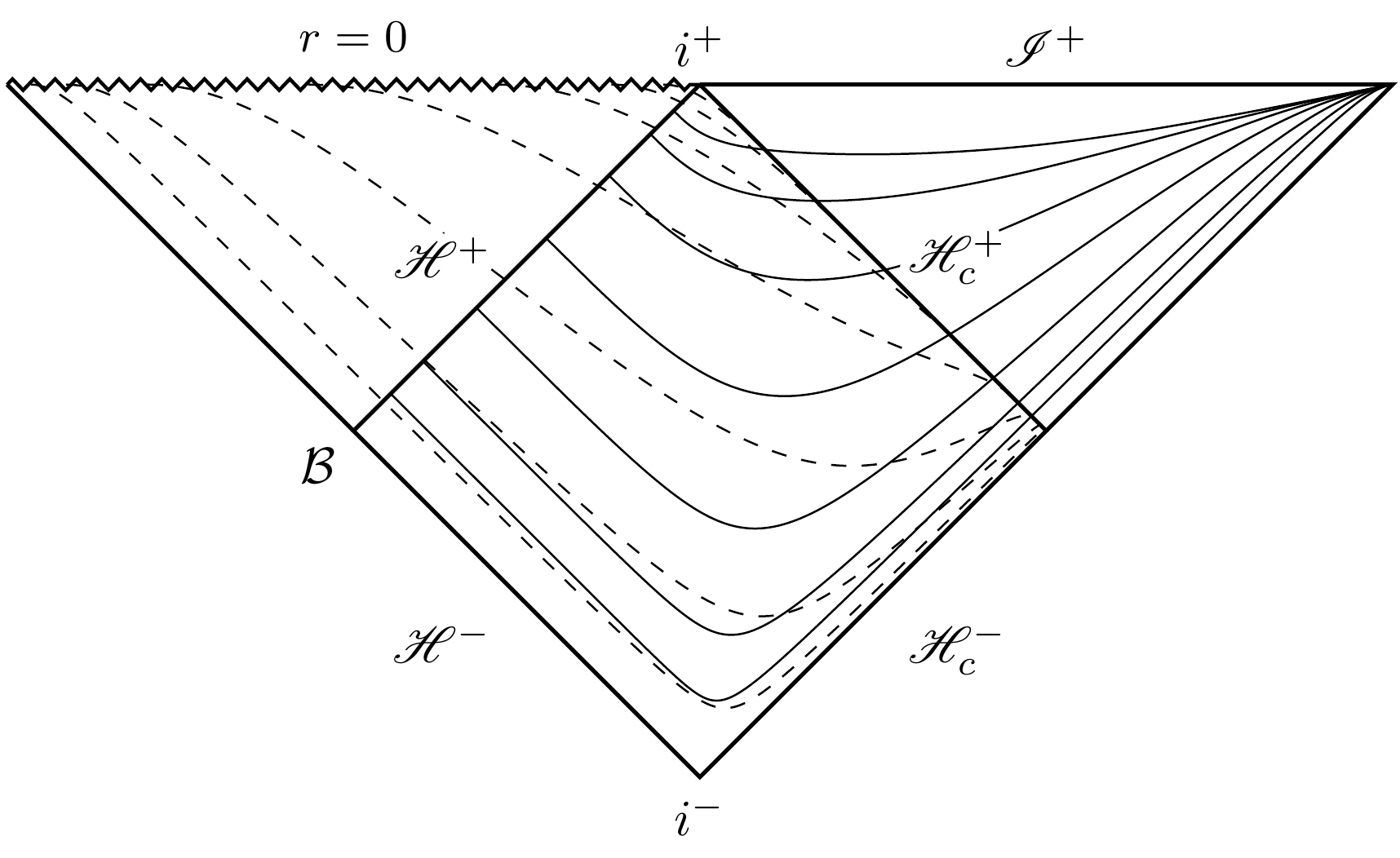}
 \end{minipage}
 \begin{minipage}[t]{0.49\textwidth}
  \centering
  \textbf{(b) Minimal gauge}\\
  \includegraphics[width=\linewidth]{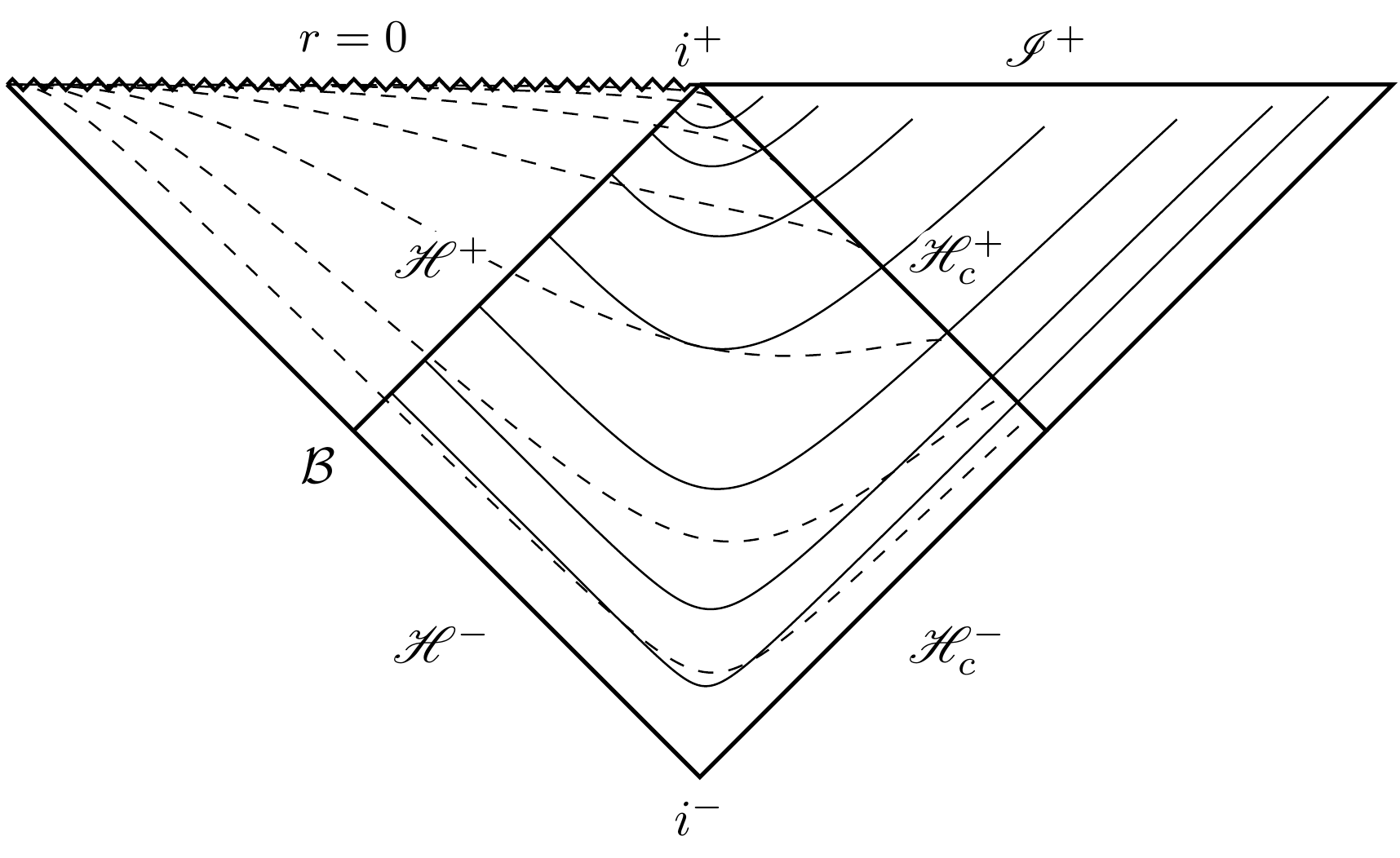}
 \end{minipage}
 \caption{Conformal diagrams of two SdS bridge families that are regular at both horizons, given by \eqref{eq:minimum_height} and \eqref{eq:minimal_height}.
 Curves show constant-time surfaces for each family.
The dashed lines use the chart regular beyond the  black-hole horizon; the solid lines use the chart regular beyond the
 cosmological horizon. The figure is taken from Fig.~11 of \cite{zenginoglu_bridge_2025}.}
 \label{fig:bridge_foliations}
\end{figure*}

Figure~\ref{fig:bridge_foliations} shows how these horizon-regular slices
continue toward the conformal boundary at finite $L$.  The minimum-height
slices meet at the asymptotic end.  The minimal-gauge slices reach distinct
cuts of the future conformal boundary.  A regular hyperboloidal flat limit
requires this boundary foliation to remain smooth as $L\to\infty$, with a
nondegenerate conformal spatial metric \cite{zenginoglu_bridge_2025,ZhouPanossoMacedo:2025}.  Below, we express this geometric requirement as an asymptotic condition on the boost.

\subsection{Compactification and the flat limit}

In addition to \eqref{eq:minimum_height} and
\eqref{eq:minimal_height}, we consider the following foliations as determined by the boost functions \cite{zenginoglu_bridge_2025}:
\begin{align}
B_{\rm lin}&=-1+2\frac{r_c-r}{r_c-r_b},
\label{eq:linear_boost}\\
B_{\rm fl}&=-1+2\frac{r_c-r}{r_c-r_b}\frac{r_b^2}{r^2},
\label{eq:flat_linear_boost}\displaybreak[1]\\
B_{\rm Mav}&=-\frac{1-3M/r}{\sqrt{1-9M^2\Lambda}}
 \sqrt{1+\frac{6M}{r}},
\label{eq:mav_boost}\\
B_{\rm sr}&=-\gamma r+\frac{\beta}{r^2};
\ \gamma=\frac{r_c^2+r_b^2}{r_c^3-r_b^3},\
\beta=\frac{r_c^2r_b^2(r_c+r_b)}{r_c^3-r_b^3}.
\label{eq:slow_roll_boost}
\end{align}
The linear boost $B_{\rm lin}$ is a simple interpolation between the required horizon values $+1$ and $-1$ from \eqref{eq:boost_conditions}.  The modified-linear boost $B_{\rm fl}$ adds the
factor $r_b^2/r^2$ to obtain a flat Schwarzschild limit to the minimal gauge as $L\to\infty$ \cite{zenginoglu_bridge_2025}.
The boost $B_{\rm Mav}$ was used in wave estimates on SdS by Mavrogiannis \cite{mavrogiannis2024quasilinear,mavrogiannis2023morawetz}.
The slow-roll boost $B_{\rm sr}$ comes from coordinates adapted to slow-roll
scalar evolution around a black hole \cite{GregoryKastorTraschen:2018},
also used in quantum-field correlation calculations
\cite{AndersonTraschen:2022,AndersonScofieldTraschen:2023}.
Its constants $\gamma$ and $\beta$ are fixed by the two horizon conditions. All these families are regular across both SdS horizons and are suitable for numerics with finite $L$. However, not all of them have a regular flat limit across future null infinity, which imposes an additional condition on their asymptotic behavior. To see this, map the horizons to fixed coordinate locations using
\begin{align}
 \rho&=\frac{1-r_b/r}{1-r_b/r_c},
 \qquad r(\rho)=\frac{r_b}{1-(1-r_b/r_c)\rho},
 \label{eq:compactification}\\
 G(r)&=\frac{d\rho}{dr}
 =\frac{r_b}{(1-r_b/r_c)r^2}.
 \label{eq:G}
\end{align}
The black-hole and cosmological horizons are at $\rho=0$ and $1$.
We use subscript $0$ to denote the Schwarzschild limit, $\Lambda\to0$
($L\to\infty$) at fixed $M$. At fixed $r$,
\begin{equation}
 \rho\to\rho_0:=1-\frac{2M}{r}.
 \label{eq:rho_limit}
\end{equation}

For the compactification \eqref{eq:compactification}, we choose the conformal factor as
\begin{equation}
 \Omega_L:=\frac{r_b}{r}
 =1-\left(1-\frac{r_b}{r_c}\right)\rho .
 \label{eq:conformal_factor}
\end{equation}
At the cosmological horizon for each finite $L$,
\begin{equation}
 \Omega_L(1)=\frac{r_b}{r_c}>0.
 \label{eq:finite_L_conformal_factor}
\end{equation}
In the Schwarzschild limit,
$\Omega_L\to\Omega_0=2M/r=1-\rho_0$, and the outer boundary becomes
$\scri^+$, where $\Omega_0=0$.

To determine which horizon-regular SdS bridges approach outgoing hyperboloidal slices in Schwarzschild, we compare their fixed-radius flat
limits and behavior at large radius.
For each family, define $B_0(r):=\lim_{L\to\infty}B(r;L)$ with $M$ and $r$ held fixed. With $x:=M/r$, the table below lists
the limits for the six families.
\begin{equation}
\begin{array}{lcc}
\text{bridge}&B_0(r)&\displaystyle\lim_{r\to\infty}B_0(r)\\[1ex]
\hline
\text{minimum height}&3x-\tfrac12&-\tfrac12\\
\text{minimal gauge}&-1+8x^2&-1\\
\text{linear}&+1&+1\\
\text{modified linear}&-1+8x^2&-1\\
\text{Mavrogiannis}&-(1-3x)\sqrt{1+6x}&-1\\
\text{slow roll}&4x^2&0
\end{array}
\label{eq:boost_limits}
\end{equation}
An outgoing Schwarzschild hyperboloidal slice must satisfy $B_0\to-1$ as
$r\to\infty$.  Regularity of its conformal spatial metric fixes the rate
of approach.  With $G_0=2M/r^2$ and $f_0=1-2M/r$, the radial coefficient is
\begin{equation}
 \bar g_{0,\rho\rho}
 =\Omega_0^2\frac{1-B_0^2}{f_0G_0^2}
 =\frac{r^2(1-B_0^2)}{f_0}.
 \label{eq:conformal_spatial_limit}
\end{equation}
A finite, positive boundary value requires
$B_0=-1+aM^2/r^2+o(M^2/r^2)$ with $a>0$, giving
$\bar g_{0,\rho\rho}\to2aM^2$.
This falloff also makes $h_0+r_{*,0}$ approach a finite constant; the
retarded time $t-r_{*,0}=\tau-(h_0+r_{*,0})$ therefore labels distinct cuts
of $\scri^+$ as $\tau$ varies.
These conditions exclude the minimum-height, linear, and slow-roll
families.  The minimal, modified-linear, and Mavrogiannis families give
regular hyperboloidal limiting slices.
We use the minimal gauge for its simple analytic form and regular
hyperboloidal Schwarzschild limit
\cite{PanossoMacedo:2023qzp,ZhouPanossoMacedo:2025}.
The Mavrogiannis foliation provides another
suitable family.

Misner's regulator uses the positive finite-$L$ boundary value
\eqref{eq:finite_L_conformal_factor} to keep the cosmological horizon at a
nonvanishing conformal factor.  To demonstrate the relevant scale,
suppose that $g$ satisfies $G_{ab}[g]+\Lambda g_{ab}=0$ and define
the conformal metric $\bar g=\Omega_L^2g$.  Then
\begin{equation}
 G_{ab}[\bar g]=\mathcal T_{ab}[\Omega_L]
 -\frac{\Lambda}{\Omega_L^2}\bar g_{ab},
 \label{eq:conformal_einstein_scaling}
\end{equation}
where $\mathcal T_{ab}$ includes the derivative terms of $\Omega_L$
\cite{Zenginoglu:2008pw}.  Evaluated at the cosmological horizon for each
finite $L$, the coefficient $\Lambda/\Omega_L^2$ approaches $3/(4M^2)$ as
$L/M\to\infty$, while the outer boundary of the compactified domain becomes Schwarzschild
future null infinity.  At fixed areal radius, by contrast, this coefficient goes to zero.

\section{Scalar waves}
\label{sec:scalar}

\subsection{Curvature coupling and first-order system}

We study the massless curvature-coupled equation
\begin{equation}
 \left(\Box_g-\xi R[g]\right)\Phi=0,
 \qquad \xi\in\left\{0,\frac16\right\},
 \label{eq:curvature_coupled_wave}
\end{equation}
where $g$ denotes the physical SdS or Schwarzschild metric.  Minimal coupling
has $\xi=0$; conformal coupling in four spacetime dimensions has $\xi=1/6$.
Write
\begin{equation}
 \Phi(t,r,\theta,\varphi)=\frac{1}{r}
 \sum_{\ell m}u_{\ell m}(t,r)Y_{\ell m}(\theta,\varphi).
 \label{eq:scalar_decomposition}
\end{equation}
Each mode satisfies
\begin{align}
 \left(-\partial_t^2+\partial_{r_*}^2-V_{\ell,\xi}\right)u=0,
 \label{eq:wave_equation}\\
 V_{\ell,\xi}=f\left[\frac{\ell(\ell+1)}{r^2}
 +\frac{f'(r)}{r}+\xi R[g]\right].
 \label{eq:scalar_potential}
\end{align}
For uniform SdS, $R[g]=12/L^2$ and therefore
\begin{equation}
 V_{\ell,\xi}=f\left[\frac{\ell(\ell+1)}{r^2}
 +\frac{2M}{r^3}+\frac{2(6\xi-1)}{L^2}\right].
 \label{eq:uniform_coupled_potential}
\end{equation}
There is a $-2/L^2$ term at $\xi=0$ which cancels at $\xi=1/6$. In Schwarzschild, $R[g]=0$, so both couplings give the same equation and the
same reference waveform.

To see the role of curvature coupling, we write the conformal transformation for the scalar wave equation.  Let
$\bar g=\Omega_L^2 g$ and $\phi=\Omega_L^{-1}\Phi$.  For the sourced wave equation,
conformal covariance gives
\begin{equation}
 \left(\Box_{\bar g}-\frac{R[\bar g]}6\right)\phi
 +\left(\frac16-\xi\right)\frac{R[g]}{\Omega_L^2}\phi
 =\Omega_L^{-3}S.
 \label{eq:conformal_source_transform}
\end{equation}
The curvature-coupling coefficient vanishes at $\xi=1/6$.  Because
$\Omega_L=r_b/r$, the evolved field is proportional to its conformal
counterpart, $\phi_{\ell m}=(u_{\ell m}/r_b)Y_{\ell m}$.

The scalar equations contain no tensorial conformal Einstein source terms,
and therefore do not fully test the regularization idea.  Nevertheless, minimal scalar coupling has a lower-order term with the same $\Lambda/\Omega_L^2$ scaling as \eqref{eq:conformal_einstein_scaling}, whereas conformal coupling cancels
it.  The two cases provide, respectively, an algebraic stress test and a
conformally regular radiative analogue.  They test whether matched scalar
observables recover their Schwarzschild counterparts in a controlled
large-$L$ limit.

Next, we discuss the first-order reduction used in the numerical implementation.  We suppress the mode labels below.  For the time evolution, define
\begin{equation}
 p:=\frac{d\rho}{dr_*}=fG, \qquad
 A:=\frac{p}{1-B^2},
 \label{eq:coefficients}
\end{equation}
and the potential coefficient
\begin{equation}
 P_{\ell,\xi}:=\frac{V_{\ell,\xi}}{p}.
 \label{eq:potential_coefficient}
\end{equation}
We write the equation in first-order symmetric hyperbolic form using auxiliary variables defined as
\begin{equation}
 \psi:=\partial_\rho u,\qquad
 \pi:=\frac{1-B^2}{p}\partial_\tau u-B\psi.
 \label{eq:first_order_variables}
\end{equation}
The evolution system becomes
\begin{align}
 \partial_\tau u&=A(B\psi+\pi),
 \label{eq:evolve_u}\\
 \partial_\tau\psi&=\partial_\rho\!\left[A(B\psi+\pi)\right],
 \label{eq:evolve_psi}\\
 \partial_\tau\pi&=\partial_\rho\!\left[A(\psi+B\pi)\right]-P_{\ell,\xi}u.
 \label{eq:evolve_pi}
\end{align}
To check the accuracy of the evolution, we monitor the constraint
\begin{equation}
 C:=\psi-\partial_\rho u.
 \label{eq:constraint}
\end{equation}
The radial characteristic speeds are
\begin{equation}
 c_-=-A(1+B)=-\frac{p}{1-B}, \quad c_+=A(1-B)=\frac{p}{1+B}.
 \label{eq:characteristic_speeds}
\end{equation}
The boundary speeds are $(c_+,c_-)=(0,c_-<0)$ at $\rho=0$ and
$(c_+,c_-)=(c_+>0,0)$ at $\rho=1$.  The characteristic flow determines all
boundary values.  For a
simple crossing with $B'(r_h)\ne0$, analytic cancellation gives
\begin{equation}
 A_h=\frac{f'(r_h)G(r_h)}{-2s_hB'(r_h)},\qquad
 s_b=+1,\quad s_c=-1.
 \label{eq:A_endpoint}
\end{equation}
We cancel every height-function pole algebraically and assign
\eqref{eq:A_endpoint} at each horizon.

The evolution domain ends at the cosmological horizon, $\hc$. No data from beyond the horizon is needed. This is similar to the excision procedure in numerical relativity, which motivates the title of Misner's paper \cite{misner_excising_2006}. We use the analytic continuation of the height function only as
a geometric diagnostic of the foliation family and of the asymptotic end it
approaches as $L/M\to\infty$.  The finite-$L$ evolutions test the regular
horizon coefficients and extract the outer waveform on $\hc$.

For the minimal gauge, the limiting Schwarzschild coefficients are
\begin{align}
 B_0&=-1+2(1-\rho_0)^2, \quad A_0=\frac{1}{8M(2-\rho_0)},
 \label{eq:schwarz_coefficients}\\
 &P_{\rm Schw}=\frac{\ell(\ell+1)+(1-\rho_0)}{2M}.
 \label{eq:schwarz_potential}
\end{align}
The Schwarzschild code implements these expressions separately from the finite-$L$ geometry.
For uniform SdS at fixed interior $\rho$, both finite-$L$ potentials approach this result.  At the cosmological horizon, however, the limits are
\begin{align}
 \lim_{L\to\infty}P_{\ell,0}(1)
 &=\frac{\ell(\ell+1)-2}{2M},\notag\\
 \lim_{L\to\infty}P_{\ell,1/6}(1)
 &=\frac{\ell(\ell+1)}{2M}=P_{\rm Schw}(1).
 \label{eq:coupling_endpoint_limits}
\end{align}

\subsection{Matched data, sources, clocks, and observables}
\label{sec:protocol}
\label{sec:localized_source}

To isolate the effect of artificial cosmology, we compare waveforms using matched data and a common time normalization.  We match the physical data and source geometrically so that changes in their coordinate representation do not introduce a different physical solution. For the data choices below, we use the peak-normalized bump
\begin{equation}
 b(x)=
 \begin{cases}
 \exp[1-1/(1-x^2)],&|x|<1,\\
 0,&|x|\geq1.
 \end{cases}
 \label{eq:bump}
\end{equation}

\paragraph{Fixed-window data.}
We first test prompt and ringdown waveform recovery on retarded-time
intervals held fixed as $L$ varies, separating finite-time agreement from
the eventual cosmological decay.
The finite-time comparison uses the same areal-radius profile on every
background,
\begin{equation}
 u(0,r)=b\!\left(\frac{r-4M}{1.5M}\right),\quad
 \psi=D_\rho u,\quad \pi=-B\psi,
\label{eq:displacement_data}
\end{equation}
with support $2.5M<r<5.5M$.  Here $D_\rho$ is the spectral differentiation
operator; $\partial_\tau u=\partial_tu=0$ initially, so the data are stationary with respect to Killing time.

\paragraph{Tail data.}
To probe weak late-time decay separately, we use an initially dynamical
pulse with zero displacement and nonzero velocity.  The Schwarzschild solution has the Price-law decay. We prescribe
\begin{equation}
 \begin{aligned}
 u&=\psi=0, &G_{\rm v}(r)&=b\!\left(\frac{r-6M}{3M}\right),\\
 \partial_\tau u&=G_{\rm v}, &\pi&=G_{\rm v}/A,
 \end{aligned}
 \label{eq:velocity_profile}
\end{equation}
with support $3M<r<9M$.  In Eqs.~\eqref{eq:displacement_data} and
\eqref{eq:velocity_profile}, the common areal profiles and prescribed
Killing-time derivatives define data on $L$-dependent slices approaching the
chosen Schwarzschild slice.

\paragraph{Retarded time.}
Matched waveforms require a common retarded-time origin.  We set $r_0=4M$
and normalize
\begin{equation}
 h_L(r_0)=0,\qquad r_{*,L}(r_0)=0.
\end{equation}
For a bridge with this normalization, the offset between $\tau$ and the
outgoing null coordinate at the cosmological horizon is
\begin{equation}
 q_B(L;r_0):=\lim_{r\to r_c}(h_L+r_{*,L})
 =\int_{r_0}^{r_c}\frac{1+B}{f}\,dr.
 \label{eq:general_q}
\end{equation}
The integrand is regular at $r_c$, where $U=\tau-q_B=t-r_{*,L}$.
For the minimal gauge, cancellation of the horizon logarithms and the
identity $(2\kappa_b)^{-1}-(2\kappa_c)^{-1}+(2\kappa_u)^{-1}=0$ give
\begin{equation}
q_L=\frac{1}{\kappa_b}
\ln\frac{r_0(r_c-r_b)}{r_c(r_0-r_b)}
\longrightarrow q_0.
\label{eq:q_L}
\end{equation}
The finite Schwarzschild limit is
\begin{equation}
 q_0=4M\ln\frac{r_0}{r_0-2M},
 \label{eq:q_zero}
\end{equation}
which reduces to $q_0=4M\ln2$ for the reference radius $r_0=4M$.
Waveforms are compared at the geometric retarded time
\begin{equation}
 U=\tau-q_L.
\label{eq:retarded_time}
\end{equation}
At $\hc$, $U$ is the outgoing retarded-time coordinate.  The same
normalization reaches Schwarzschild $\scri^+$ in the limit.  At finite
radius, $U$ is a stationary clock normalized at the outer boundary.  Its
relation to the local outgoing coordinate is
\begin{equation}
 U-(t-r_{*,L})=h_L(r)+r_{*,L}(r)-q_L.
\label{eq:finite_radius_clock}
\end{equation}
An additive clock offset preserves the asymptotic decay exponent but changes
the finite-time logarithmic slope used to estimate it.  We therefore use the
common geometric time origin in \eqref{eq:retarded_time} when comparing local
power indices and transition times.

\paragraph{Localized source.}
\label{sec:localized_protocol}

To test artificial cosmology beyond single-mode radial evolution, we consider
a spatially localized scalar response with nontrivial angular structure.
This tests Schwarzschild recovery for multidimensional wave propagation,
including focusing and caustic echoes
\cite{DolanOttewill:2011,ZenginogluGalley:2012,CasalsNolan:2023}.
We exploit the spherical background to evolve independent radial modes and
reconstruct the three-dimensional field.  We use a normalized source localized
in radius, time, and angle as a smooth approximation to a Dirac delta function.  The resulting field is the retarded Green function integrated
against this source profile.
We compare the full localized response across backgrounds.  With zero initial data,
\begin{equation}
 \left(\Box_g-\xi R[g]\right)\Phi=S,
 \qquad \Phi|_{\tau=0}=\partial_\tau\Phi|_{\tau=0}=0.
 \label{eq:sourced_wave_equation}
\end{equation}
The same physical source $S$ is used on every background; its conformal
representation is $\Omega_L^{-3}S$.  Let $\gamma$ be the angle from the
source direction $(\theta_s,\varphi_s)$ and set
\begin{equation}
 S=\mathcal A\,T(t)\,R(r)\,\mathcal P_\kappa(\gamma),
 \label{eq:source_product}
\end{equation}
\begin{align}
 T(t)&=\frac{b\left[(t-t_s)/\sigma_t\right]}{\sigma_t\beta_0},
 &
 R(r)&=\frac{b\left[(r-r_s)/\sigma_r\right]}{\beta_2},
 \notag\\
 \mathcal P_\kappa(\gamma)&=\frac{\kappa\,e^{\kappa(\cos\gamma-1)}}
 {2\pi\left(1-e^{-2\kappa}\right)}.
 \label{eq:source_angular}
\end{align}
Here $\beta_0=\int_{-1}^{1}b(x)dx$ and
$\beta_2=\int b[(r-r_s)/\sigma_r]r^2dr$, so each factor has unit integral and
$\int\sqrt{-g}S\,d^4x=\mathcal A$. The parameter choices are given in Sec.~\ref{sec:localized_results}.

The real-spherical-harmonic coefficients are
\begin{equation}
 a_{\ell m}=\frac{i_\ell(\kappa)}{i_0(\kappa)}
 Y_{\ell m}(\theta_s,\varphi_s),
 \label{eq:source_modal_amplitudes}
\end{equation}
where $i_\ell$ is a modified spherical Bessel function.  The $m$ dependence
factors out, so one response per retained $\ell$ solves
\begin{equation}
 \left(-\partial_t^2+\partial_{r_*}^2-V_{\ell,\xi}\right)v_\ell
 =f r\,\mathcal A T(t)R(r).
 \label{eq:sourced_radial_equation}
\end{equation}
In the first-order system the source is evaluated explicitly at the Killing
time $t=\tau-h_L(r)$, ensuring the same emitter on every background.  With
$u_{\ell m}=a_{\ell m}v_\ell$, the reconstructed waveform is
\begin{equation}
 u(U,\theta,\varphi)
 =\sum_{\ell m}a_{\ell m}v_\ell(U)Y_{\ell m}(\theta,\varphi),
 \label{eq:angular_reconstruction}
\end{equation}
at the extraction radius, using the common retarded time
\eqref{eq:retarded_time}.

At the outer boundary, the $\xi=0$ observable is the
reduced field $W_L(U)\equiv W_{L,0}(U):=u_2(U,\rho=1)$ on $\hc$; its
Schwarzschild counterpart $W_{\rm Schw}(U):=u_2(U,\rho_0=1)$ is evaluated at
$\scri^+$.  Thus $W$ is the
relevant spherical-harmonic coefficient of the radiation-scaled field
$r\Phi$.  For a common time interval
$\mathcal J$, the direct fixed-mode waveform error is
\begin{equation}
 E_2(L;\mathcal J)=
 \frac{\|W_L-W_{\rm Schw}\|_{L^2(\mathcal J)}}
      {\|W_{\rm Schw}\|_{L^2(\mathcal J)}},
 \label{eq:regulator_E2}
\end{equation}
where $W_{\rm Schw}$ is computed with the separate Schwarzschild implementation.  For the localized source, $W$ denotes the
outer-boundary fields $\{u_{\ell m}\}$ and the primary comparison uses the
sphere-integrated Parseval norm
\begin{equation}
 \|W\|_{\mathcal J,S^2}^2=
 \int_{\mathcal J}dU\sum_{\ell m}|u_{\ell m}(U)|^2.
 \label{eq:parseval_norm}
\end{equation}
The corresponding relative error is
\begin{equation}
 E_{2,S^2}(L;\mathcal J)=
 \frac{\|W_L-W_{\rm Schw}\|_{\mathcal J,S^2}}
      {\|W_{\rm Schw}\|_{\mathcal J,S^2}}.
 \label{eq:localized_regulator_E2}
\end{equation}
Besides comparing individual finite-$L$ waveforms with Schwarzschild, we
test whether combining several cosmological lengths improves the recovery.
For minimal coupling, we model the large-$L$ dependence at each common
retarded time $U$ as
\begin{equation}
 W_{L,0}=W_{\rm Schw}+c_{1,0}\frac{M}{L}
 +c_{2,0}\left(\frac{M}{L}\right)^2
 +\mathcal O\!\left((M/L)^3\right).
 \label{eq:regulator_expansion}
\end{equation}
At each $U$, the coefficients $c_{1,0}$ and $c_{2,0}$ are independent of $L$. Waveforms at cosmological lengths $L$, $2L$, and $4L$ then give the
extrapolated estimate
\begin{equation}
 W_\infty^{(L)}=
 \frac{W_L-6W_{2L}+8W_{4L}}{3},
 \label{eq:nested_extrapolant}
\end{equation}
which cancels the terms proportional to $M/L$ and $(M/L)^2$ without using
the Schwarzschild waveform as input.  Repeating the construction with
$(2L,4L,8L)$ gives a second, overlapping extrapolant.  We compare the two
estimates with each other and with the separately evolved Schwarzschild
control to test the extrapolation. A conformally coupled sequence
requires a separate test of its finite-$L$ expansion.

The fixed-mode comparison uses cumulative intervals that begin at $U=0$ and
end at $U/M=40$, $80$, and $160$, together with the disjoint intervals
$40\leq U/M\leq80$ and $80\leq U/M\leq160$.  The localized-source comparison
uses the common simulated interval, from the first nonnegative output time
through the last time reached by every evolution.
The uniform waveform and localized-source sequences use linear interpolation
onto a common $U$ grid; the exterior ringdown and matched-tail comparisons
use cubic interpolation.

\subsection{Numerical methods and error control}
\label{sec:numerical-methods}

We tested artificial cosmology using different numerical methods. The uniform waveform sequence and the large-$L$ tail benchmark
use Dedalus~3 \cite{burns_dedalus_2020} with a Chebyshev-$T$ basis and RK222.
The localized source uses eighth-order finite differences on a uniform
radial grid and classical RK4, with source evaluation at each stage.
The exterior ringdown and matched-tail comparisons use continuous
Legendre--Gauss--Lobatto spectral elements \cite{Gassner:2013SBP} and
third-order Radau IIA \cite{HairerWanner:1999Radau}, evolving $u$ and
$\partial_\tau u$ directly. All calculations evolve the characteristic boundaries without prescribed boundary data and use the geometric clock
normalization defined above.

Each simulation sequence is tested at three resolutions.  Uniform and
localized-source comparisons use matched Schwarzschild controls.  We distinguish
discretization error from sensitivity to the estimator, comparison window,
and sampling interval, and
from physical source-width dependence.  Appendix~\ref{app:convergence} gives
the resolution sequences and independent checks;
Appendix~\ref{app:latetime} gives the tail-specific criteria and convergence
results.

\section{Recovering Schwarzschild solutions with artificial cosmology}
\label{sec:results}

In this section, we quantify the waveform error when a cosmological horizon
replaces null infinity at finite $L$.
The same comparisons also test the isolated-system approximation in a
cosmological spacetime, motivated by the small positive cosmological constant.  Increasing $L/M$, we study how scalar waveforms approach their asymptotically flat limits over finite observation intervals. We also check different aspects of the regulator, including the convergence of extrapolated waveforms and the recovery of a localized multimode response.  The results show that the Misner regulator is effective for both minimal and conformal coupling, and that the finite-$L$ error can be reduced to a few percent with $L/M\sim 10^2$.

\subsection{Direct and extrapolated fixed-mode recovery}
\label{sec:direct_recovery}
\label{sec:extrapolated_recovery}

\begin{figure}[t]
 \centering
 \includegraphics[width=1.05\columnwidth]{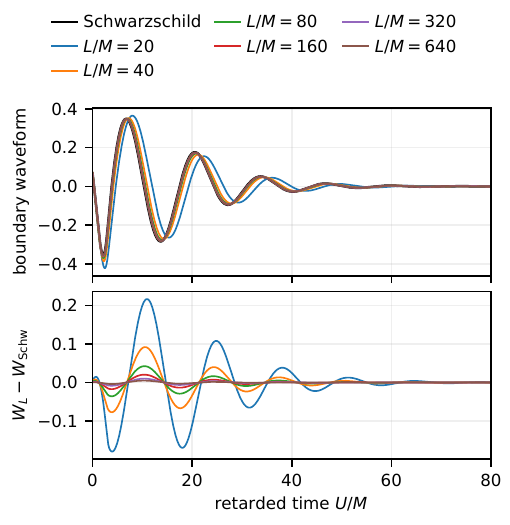}
 \caption{Minimally coupled ($\xi=0$), pure-$\ell=2$ sequence at
 the outer boundary.  Upper
 panel: waveforms for $L/M=20$ to $640$ against the separately evolved
 Schwarzschild reference.  Lower panel: the residual $W_L-W_{\rm Schw}$ on
 the same retarded time $U=\tau-q_L$, with no other fitting or rescaling.  The residual amplitude falls with $L$. Compare with Table \ref{tab:direct-regulator-errors}.}
 \label{fig:flat_sequence}
\end{figure}

Figure~\ref{fig:flat_sequence} shows how the boundary waveforms approach
the Schwarzschild solution as $L/M$ increases from $20$ to $640$.  The
prompt-pulse and ringdown structure is maintained throughout the sequence.
In the large-$L$ regime, the dominant residual amplitude falls by about a
factor of two with each doubling of $L$.

Table~\ref{tab:direct-regulator-errors} quantifies this convergence and gives
the benchmark for the regulator: the direct waveform error $E_2$ against the separately
evolved Schwarzschild solution, measured over the same interval
$0\leq U/M\leq80$ for every $L$.  The error decreases from $63.96\%$ at
$L/M=20$ to $1.34\%$ at $L/M=640$.  The numerical medium-to-fine changes
are much smaller, showing that the finite-$L$ difference significantly dominates over discretization error.

\begin{table}[t]
 \caption{Direct waveform error for the minimally coupled pure-$\ell=2$
 sequence on $0\leq U/M\leq80$.  $E_2$ is the fine-grid discrepancy from
 Schwarzschild [Eq.~\eqref{eq:regulator_E2}]; $\delta_{mf}$ is the numerical
 medium-to-fine change of the paired residual
 [Eq.~\eqref{eq:paired_refinement_change}].  Both columns are percentages
 of the Schwarzschild norm.}
 \label{tab:direct-regulator-errors}
 \begin{ruledtabular}
 \begin{tabular}{rrr}
 $L/M$ & $E_2$ (\%) & $\delta_{mf}$ (\%)\\
 \colrule
 20  & 63.96 & 0.033\\
 40  & 26.08 & 0.094\\
 80  & 11.75 & 0.069\\
 160 & 5.57  & 0.041\\
 320 & 2.72  & 0.064\\
 640 & 1.34  & 0.024\\
 \end{tabular}
 \end{ruledtabular}
\end{table}

For these data and this norm, the first tested lengths to meet $5\%$ and
$2\%$ tolerances are $L/M=320$ and $640$, respectively. 
The approximate $M/L$ falloff is encoded in the extrapolation model
\eqref{eq:regulator_expansion}.
At fixed outgoing retarded time $\widehat U=t-r_{*,0}$, a smooth
Schwarzschild field can have the expansion
$u=F(\widehat U)+G(\widehat U)/r+\mathcal O(r^{-2})$, with
$2\,dG/d\widehat U=\ell(\ell+1)F$.  Since $r_c\sim L$, such corrections
show how an $\mathcal O(M/L)$ boundary discrepancy can coexist with
an $\mathcal O(L^{-2})$ local metric deformation.

To test sensitivity to higher-order finite-$L$ corrections, we use extrapolation and combine triples $(80,160,320)$ and
$(160,320,640)$ in units of $L/M$. The largest fine-grid residual among
$W_\infty^{(80)}$, $W_\infty^{(160)}$, and the separately evolved
Schwarzschild waveform is $0.0313\%$.  The largest numerical medium-to-fine change
among these residuals is $0.409\%$, and
the largest propagated Richardson estimate is $0.119\%$. We can conclude that the extrapolated waveforms agree with the separately evolved Schwarzschild waveform at $1\%$. 
Appendix~\ref{app:convergence} gives
the numerical refinement and extrapolation checks.

\subsection{Localized multimode response}
\label{sec:localized_results}

To extend the $\xi=0$ benchmark beyond a single spherical harmonic, we solve for the localized source family described in Section~\ref{sec:localized_source}.  We use $r_s=6M$, $\sigma_r=0.75M$, $t_s=30M$,
$\sigma_t=2M$, $\kappa=64$, and $(\theta_s,\varphi_s)=(\pi/2,0)$.
These choices give radial support $5.25M<r<6.75M$, temporal support
$28M<t<32M$, and angular scale $\kappa^{-1/2}=0.125$ radians.  Modes with
$|a_{\ell m}|$ below $10^{-13}$ of the maximum are omitted.  We extract
waveforms at $r=8M$, $r=12M$, and the outer boundary; the comparison
uses the outer-boundary field on the common interval
$0\lesssim U/M\leq57.2$, before the late-time signal.

\begin{figure}[t]
 \centering
 \includegraphics[width=\columnwidth]{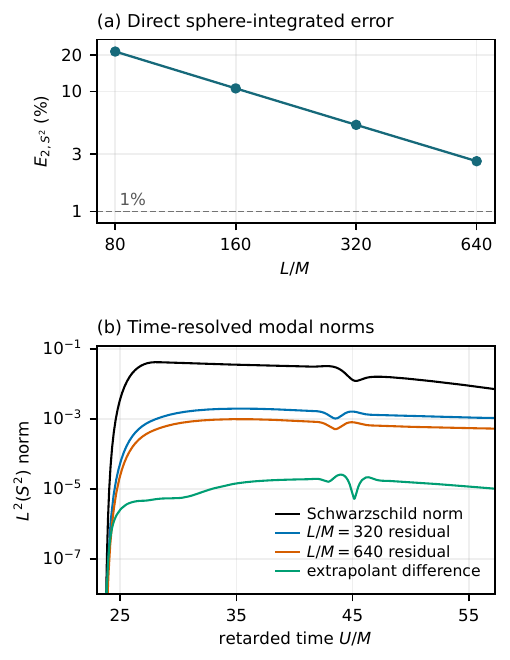}
 \caption{Minimally coupled localized source on the common simulated
 interval.  (a) Direct sphere-integrated errors $E_{2,S^2}$ in percent;
 the dashed line marks $1\%$.  (b) Instantaneous absolute $L^2(S^2)$ norms
 of the Schwarzschild field, the
 finite-$L$ residuals against Schwarzschild, and the difference between
 the two nested extrapolants.  The norms include all retained modes and
 integrate over the entire observation sphere.}
 \label{fig:localized}
\end{figure}

In the sphere-integrated norm \eqref{eq:parseval_norm}, the direct differences are
$E_{2,S^2}(320)=5.27\%$ and $E_{2,S^2}(640)=2.63\%$.  The
$W_\infty^{(80)}$ and $W_\infty^{(160)}$ extrapolants differ by $0.0517\%$;
their Schwarzschild differences are $0.0584\%$ and
$0.00672\%$, respectively. Figure~\ref{fig:localized} shows how the error decreases with $L$ and how the residuals vary over the signal.

\subsection{Large-\texorpdfstring{$L$}{L} Price-law benchmark}
\label{sec:uniform-price-benchmark}

Tails provide a stricter test than the prompt waveform. To test Schwarzschild decay-rate at the asymptotic boundary with artificial cosmology, we evolve matched $\ell=1$, $\xi=0$ data at $L/M=3072$, extracting the SdS waveform at $\hc$, and we compare to an independently evolved Schwarzschild reference at $\scri^+$.

For each extracted waveform $W(U)$, we define the centered root-mean-square
envelope
\begin{equation}
 A(U)=\left[\frac{1}{\Delta U}
 \int_{U-\Delta U/2}^{U+\Delta U/2}|W(s)|^2\,ds\right]^{1/2},
 \label{eq:rms-envelope}
\end{equation}
with $\Delta U=10M$, evaluated as an average of squared waveform samples.
We estimate the local power index $p_{\rm eff}=-d\ln A/d\ln U$ by fitting
$\ln A$ against $\ln U$ over a centered $40M$ window.
For a slowly varying power-law tail, $A\simeq|W|$, so this index approximates
the decay exponent.

The SdS local power index remains within $5\%$ of $p=3$ on
$232\lesssim U/M\lesssim382$.
Spatial and timestep checks give a spread of about $2M$ (see Appendix~\ref{app:large-L-tail} for details).  These checks indicate that an intermediate interval is compatible with the Schwarzschild Price-law rate. 

\begin{figure}[tbp]
 \centering
 \includegraphics[width=\columnwidth]{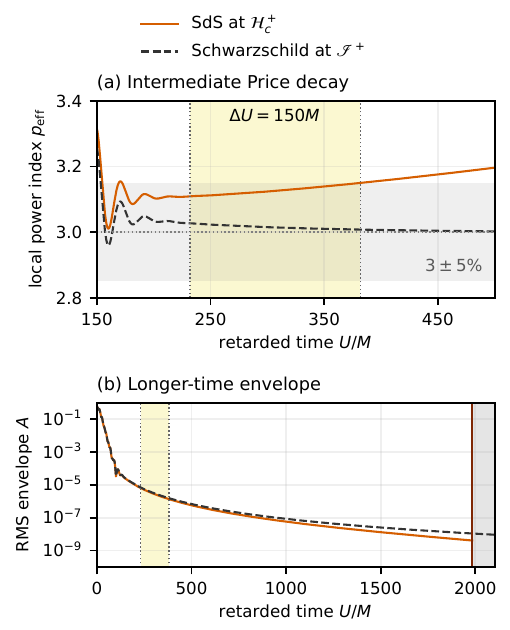}
 \caption{Intermediate, polynomial tail decay rate in minimally coupled
 $\ell=1$ SdS at $L/M=3072$.  Upper panel: the local power indices at
 $\hc$ (solid orange) and Schwarzschild $\scri^+$ (dashed black). The polynomial rate is compatible within the range $p=3\pm5\%$ in the interval $232\lesssim U/M\lesssim382$. The decay slightly is stronger in SdS. Lower panel: the centered $10M$
 root-mean-square envelopes defined in \eqref{eq:rms-envelope}. We see that the SdS waveform is slightly smaller than the Schwarzschild one, but the difference is small and consistent with the expected cosmological expansion.}
 \label{fig:large-L-tail}
\end{figure}

Figure~\ref{fig:large-L-tail} shows the Price-law interval within
the longer waveform.  The common evolution extends to $U/M\simeq1984$, or $\kappa_cU=0.645$. In the experiments that demonstrate the polynomial decay, the cosmological exponential regime is not reached before the numerical floor. This is reasonable and supports the well-known idea that the isolated system approximation is valid for long times before cosmological effects become relevant. Appendix \ref{app:large-L-tail} gives further details.

\subsection{Cosmological decay}
\label{sec:sds_physics}

At smaller $L$, the exponential cosmological decay becomes accessible within
the resolved evolution.  A separate minimally coupled $\ell=1$ study at
$r=8M$ measures both departure from the Schwarzschild decay rate and entry
into the cosmological regime.  For $L/M=80$ and $160$, the median departure
times are $U=115.9M$ and $133.5M$, respectively, whereas cosmological decay
begins at $\kappa_cU=2.83$ and $2.94$.  Doubling $L$ therefore changes the
departure time only modestly in black-hole units, while entry occurs at
nearly the same time in cosmological units.  These two cases provide
qualitative evidence for distinct timescales.

Both departures precede a settled Schwarzschild Price tail: at the same
radius, the reference decay index does not remain within $5\%$ of its
$\ell=1$ Price-law value until $U=220.3M$.  These runs resolve a transition
from a Schwarzschild-like transient to cosmological decay.  The large-$L$
boundary test above resolves an intermediate Price-law interval; its later
cosmological regime remains numerically unresolved.
Appendix~\ref{app:finite-radius-crossover} gives the transition criteria
and their sensitivity to the decay-rate estimator.

\section{Exterior-supported artificial cosmology}
\label{sec:exterior-supported}

Uniform SdS alters the metric throughout the static patch, including the
strong-field region. But Misner's proposal primarily concerns the regularization of the asymptotic domain. Therefore, it makes sense to confine the cosmological source to an exterior
region while retaining the same asymptotic value
\begin{equation}
 \Lambda=\frac{3}{L^2}.
 \label{eq:exterior_lambda}
\end{equation}
We would like the cosmological source to vanish in a domain containing the event horizon and the photon sphere. Then, in an interior domain, the metric, boost, and scalar potential are exactly Schwarzschild. However, the outer layer, cosmological boundary, and clock still affect the waveform.
We perform comparisons with uniform SdS at fixed $L$ to test whether this construction
reduces the total waveform error. We first construct the source and its metric based on an idea by Dymnikova \cite{Dymnikova:2000algebraic,Dymnikova:2002mass} relying on the static spherical ansatz. We then specify
the foliation and compare ringdown and tail signals.  

\subsection{Conserved cosmological tensor and exact metric}

Restricting the cosmological term to an exterior region is not immediately trivial. The term $\Lambda$ is called cosmological \emph{constant} for a geometric reason. Simply replacing $\Lambda$ by a radial profile $\lambda(r)$ in the vacuum
equations violates the contracted Bianchi identity unless it is constant:
\begin{equation}
 G_{ab}+\lambda g_{ab}=0
 \quad\Longrightarrow\quad
 0=\nabla^aG_{ab}=-\nabla_b\lambda.
 \label{eq:variable_lambda_bianchi}
\end{equation}
But we can replace the cosmological term by a conserved \emph{tensor},
\begin{equation}
 G^a{}_b+\boldsymbol{\Lambda}^a{}_b=0,
 \qquad \nabla_a\boldsymbol{\Lambda}^a{}_b=0.
 \label{eq:cosmological_tensor_equations}
\end{equation}
In static spherical symmetry, radial variation is compatible with
conservation if the source includes suitable anisotropic stresses.
Following Dymnikova's anisotropic-vacuum framework
\cite{Dymnikova:2000algebraic,Dymnikova:2002mass}, we choose equal temporal
and radial eigenvalues $\Lambda_{\parallel}$ and angular eigenvalues
$\Lambda_{\perp}$:
\begin{equation}
 \boldsymbol{\Lambda}^a{}_b
 =\operatorname{diag}\!\left(
 \Lambda_{\parallel},\Lambda_{\parallel},
 \Lambda_{\perp},\Lambda_{\perp}\right).
 \label{eq:exterior_cosmological_tensor}
\end{equation}
For an arbitrary smooth profile $\lambda(r)$, set
\begin{align}
 \Lambda_{\parallel}(r)&=\lambda+\frac{r}{3}\lambda',
 \label{eq:lambda_parallel}\\
 \Lambda_{\perp}(r)&=\lambda+r\lambda'
 +\frac{r^2}{6}\lambda''.
 \label{eq:lambda_perp}
\end{align}
Equality of the temporal and radial eigenvalues cancels the metric-derivative
terms in the radial conservation equation, leaving
\begin{equation}
 \nabla_a\boldsymbol{\Lambda}^a{}_r
 =\Lambda_{\parallel}'
 +\frac{2}{r}\left(\Lambda_{\parallel}-\Lambda_{\perp}\right)=0.
 \label{eq:exterior_conservation_check}
\end{equation}
The contributions proportional to $\lambda'$ and $\lambda''$ cancel exactly
for any smooth profile; symmetry makes the other components vanish.
Where $\lambda$ is constant, the tensor reduces to
$\lambda\delta^a{}_b$.  Where it varies, the derivative terms supply the
additional stresses required by conservation.  Thus $\lambda(r)$
parametrizes a conserved anisotropic source, allowing us to circumvent the obstacle against  a variable vacuum term $\lambda(r)g_{ab}$.

We now choose the profile to give a Schwarzschild interior and an SdS
exterior.  Set
$\lambda(r)=\Lambda\chi_L(r)$, with $\chi_L=0$ below $R_0$ and $\chi_L=1$
above $R_1$.  To join these regions smoothly, let $\eta(x)=e^{-1/x}$ for
$x>0$ and zero otherwise, and define
$S(x)=\eta(x)/[\eta(x)+\eta(1-x)]$.  For $2M<R_0<R_1<r_c$, the switch is
\begin{equation}
 \chi_L(r)=
 \begin{cases}
 0,&r\leq R_0,\\[1mm]
 S\!\left(\zeta_L(r)\right),&R_0<r<R_1,\\[1mm]
 1,&r\geq R_1,
 \end{cases}
 \label{eq:exterior_switch}
\end{equation}
where $\zeta_L$ is a smooth monotone map with
$\zeta_L(R_0)=0$ and $\zeta_L(R_1)=1$.
We choose the transition radii so that both the transition layer and the
outer exact-SdS region retain nonzero
widths in the compactified radial coordinate as $L\to\infty$.
Appendix~\ref{app:exterior-implementation} gives the explicit profile.

The Einstein equations with this source admit the metric
\begin{equation}
 ds^2=-f_\chi(r)dt^2+\frac{dr^2}{f_\chi(r)}+r^2d\omega^2,
 \label{eq:exterior_metric}
\end{equation}
with
\begin{equation}
 \begin{aligned}
 f_\chi(r)&=1-\frac{2M}{r}-\frac{r^2}{3}\lambda(r)\\
 &=1-\frac{2M}{r}-\frac{r^2}{L^2}\chi_L(r).
 \end{aligned}
 \label{eq:exterior_f}
\end{equation}
The integration constant $M$ is the same mass parameter in the interior and
exterior.  For $M>0$, the metric is exactly Schwarzschild below $R_0$ and SdS
above $R_1$.  Between them, the derivative terms in the source support a
smooth, nonvacuum transition layer.

The enclosed Misner--Sharp mass is
\begin{equation}
 f_\chi=1-\frac{2m_\chi}{r},
 \qquad
 m_\chi(r)=M+\frac{r^3\lambda(r)}{6}.
 \label{eq:exterior_mass_function}
\end{equation}
It includes the cosmological source as well as the central mass parameter
$M$, which remains unchanged.

Interpreting the source as a stress tensor gives
$T^a{}_b=-(8\pi)^{-1}\boldsymbol{\Lambda}^a{}_b
=\operatorname{diag}(-\varepsilon,p_r,p_\perp,p_\perp)$, with
$\varepsilon=\Lambda_{\parallel}/(8\pi)$,
$p_r=-\varepsilon$, and $p_\perp=-\Lambda_{\perp}/(8\pi)$.
The density $\varepsilon=(\lambda+r\lambda'/3)/(8\pi)$ obeys
$m_\chi'=4\pi r^2\varepsilon$.  The energy-condition combinations are
\begin{equation}
 \varepsilon+p_r=0,
 \qquad
 \varepsilon+p_\perp=-\frac{r}{16\pi}\Lambda_{\parallel}'.
 \label{eq:exterior_nec}
\end{equation}
The radial null energy condition is satisfied.  Since
$\int_{R_0}^{R_1}\Lambda_{\parallel}'dr=\Lambda>0$, the tangential condition
is violated somewhere in the layer.  The regulator therefore cannot be
interpreted as ordinary matter satisfying the null energy condition.

Extending the construction to nonlinear Einstein evolution would require
an evolution law for the source, beyond background conservation.  A metric
perturbation changes the
connection, so source perturbations must also satisfy
\begin{equation}
 \delta\!\left(\nabla_a\boldsymbol{\Lambda}^a{}_b\right)=0.
 \label{eq:perturbed_conservation}
\end{equation}
The background tensor therefore cannot, in general, be held fixed during
nonlinear evolution.  Such an extension needs a covariant conserved regulator.

The Ricci scalar quantifies the curvature introduced by the source:
\begin{equation}
 R_\chi=4\lambda+\frac{8}{3}r\lambda'
 +\frac{1}{3}r^2\lambda''.
 \label{eq:exterior_ricci_scalar}
\end{equation}
It vanishes in the Schwarzschild interior and equals $4\Lambda$ in the SdS
exterior.  The Misner--Sharp mass accumulated across the smooth layer is
\begin{equation}
 \frac{\Delta m_{\rm tr}}{M}
 :=\frac{m_\chi(R_1)-m_\chi(R_0)}{M}
 =\frac12\frac{(R_1/M)^3}{(L/M)^2}.
 \label{eq:transition_mass}
\end{equation}
Appendix~\ref{app:exterior-implementation} gives the large-$L$ scaling of
the curvature and accumulated mass.
At $L/M=640$, the layer ends at $R_1\simeq440.4M$ and contains
$\Delta m_{\rm tr}=104.3M$.  Uniform SdS has
$m_{\rm SdS}(r)=M+r^3/(2L^2)$, so both constructions have exactly the same
enclosed mass at $R_1$ and beyond.  The derivative contribution to the
source density compensates for removing the cosmological source from the
interior, while retaining the same exterior mass parameter.  
The outer geometry can differ significantly from Schwarzschild:
$|f_\chi(R_1)-f_{\rm Schw}(R_1)|=(R_1/L)^2\simeq0.474$ in this example.
This is compatible with the fixed-radius Schwarzschild limit; at finite
$L$, neither construction need approximate Schwarzschild closely at radii
comparable to $L$.  The effect on extracted radiation must instead be
assessed through waveform comparisons, as in the following subsections.

\subsection{Bridge foliation, compactification, and clock}

The exterior metric needs a bridge that crosses both horizons and recovers
the Schwarzschild foliation as $L\to\infty$.  For $M>0$, we choose
\begin{equation}
 B_\chi(r)=-1+\frac{8M^2}{r^2}
 -\frac{8M^2}{r_c^2}\chi_L(r).
 \label{eq:exterior_bridge}
\end{equation}
It has $B_\chi(2M)=1$, $B_\chi(r_c)=-1$, and $-1<B_\chi<1$ between the
horizons; beyond $r_c$, $B_\chi<-1$.  The height function satisfies
$h_\chi'=B_\chi/f_\chi$; the horizon crossings are regular, as verified in
Appendix~\ref{app:exterior-implementation}.  Since $0\leq\chi_L\leq1$ and
$r_c/L\to1$, the cosmological correction to the boost vanishes as
$\mathcal O(M^2/L^2)$.  Hence, at fixed $r$,
\begin{equation}
 B_\chi(r)=-1+\frac{8M^2}{r^2}
 \qquad(L\to\infty),
 \label{eq:exterior_bridge_flat_limit}
\end{equation}
which is the minimal-gauge Schwarzschild boost.  Likewise
$r^2\chi_L/L^2\to0$, so the
metric and bridge recover Schwarzschild at fixed radius even if $\chi_L$ has
a nonzero limiting profile.

The compact radial coordinate maps the two horizons to fixed boundary points:
\begin{equation}
 D_L=1-\frac{2M}{r_c},
 \qquad
 \rho_\chi(r)=\frac{1-2M/r}{D_L}.
 \label{eq:exterior_compactification}
\end{equation}
It places $2M$ and $r_c$ at $\rho_\chi=0$ and $1$ and goes to
$1-2M/r$ at fixed radius.  The conformal factor is
\begin{equation}
 \Omega_{\chi,L}=\frac{2M}{r},
 \qquad \Omega_{\chi,L}(1)=\frac{2M}{r_c}>0.
 \label{eq:exterior_conformal_factor}
\end{equation}
Its value at the outer boundary is positive at finite $L$ and vanishes in the
Schwarzschild limit. The coefficients multiplying spatial derivatives at the outer boundary
approach their Schwarzschild values as $L\to\infty$
(Appendix~\ref{app:exterior-implementation}).  

To compare waveforms on the same geometric clock, we keep the normalization
$r_0=4M$ and define
\begin{equation}
 q_\chi=\int_{r_0}^{r_c}\frac{1+B_\chi}{f_\chi}\,dr,
 \qquad U=\tau-q_\chi.
 \label{eq:exterior_clock}
\end{equation}
The integrand has a finite horizon limit.  Comparing its large-radius
behavior with the Schwarzschild expression gives
\begin{equation}
 q_\chi=q_0-\frac{8M^2}{r_c}
 +o\!\left(\frac{M^2}{r_c}\right)
 \Rightarrow q_0=4M\ln\frac{r_0}{r_0-2M}.
 \label{eq:exterior_clock_flat_limit}
\end{equation}
Therefore, the exterior family approaches the same Schwarzschild clock as the
uniform construction.

\subsection{Ringdown comparison}
\label{sec:exterior-tests}

We test whether preserving the near-zone geometry
improves the waveform at the outer boundary.  We compare minimally coupled
($\xi=0$), compact $\ell=2$ data at $L/M=320,640$, using matched initial
data and clocks for Schwarzschild, uniform SdS, and exterior-supported SdS.
For either cosmological background, the relative waveform error against the
same separately evolved Schwarzschild reference is
\begin{equation}
 E_X(L;\mathcal J)
 =\frac{\|W_X-W_{\rm Schw}\|_{\mathcal J}}{\|W_{\rm Schw}\|_{\mathcal J}}.
 \label{eq:exterior_waveform_error}
\end{equation}
Here $X\in\{\mathrm{SdS},\chi\}$.  We compare waveforms at the outer boundary
over the ringdown interval $15\leq U/M\leq45$.

\begin{table}[t]
\caption{Ringdown waveform errors relative to Schwarzschild for uniform
 and exterior-supported SdS, using minimally coupled ($\xi=0$), $\ell=2$
 data on $15\leq U/M\leq45$.  The relative error reduction is
 $100(1-E_\chi/E_{\rm SdS})\%$.}
\label{tab:exterior-qnm-improvement}
\begin{ruledtabular}
\begin{tabular}{rrrr}
$L/M$ & uniform SdS & exterior SdS & reduction\\
 & error (\%) & error (\%) & (\%)\\
\hline
320 & 3.18 & 2.77 & 12.8\\
640 & 1.55 & 1.42 &  8.6\\
\end{tabular}
\end{ruledtabular}
\end{table}

Restricting the cosmological source to the exterior reduces the ringdown
waveform error by $12.8\%$ at $L/M=320$ and $8.6\%$ at $L/M=640$
(Table~\ref{tab:exterior-qnm-improvement}).  

This fixed-mode test quantifies ringdown waveform agreement for the specified
initial data.
Determining the global quasinormal spectrum requires a separate spectral analysis.
Distant perturbations can strongly alter that spectrum
\cite{CheungEtAl:2022elephant} while leaving early ringdown nearly unchanged,
as demonstrated for a black hole surrounded by a dynamical thin shell
\cite{LaeugerEtAl:2025shell}.

\subsection{Exterior-supported tail comparison}
\label{sec:matched-tail-result}

Tail recovery tests the construction beyond the ringdown window.
Matched $\ell=1$ evolutions at
$L/M=640$ through $U/M=1000$ compare uniform and exterior-supported SdS at
$\hc$ with Schwarzschild at $\scri^+$, for both $\xi=0$ and $1/6$.  Here we
require the local decay index to remain within $10\%$ of $p=3$ for at least
$40M$.  Uniform conformal SdS meets this criterion; uniform minimal coupling
does not.
At fixed $L$, the field equation therefore matters for the intermediate polynomial decay rates.

\begin{figure}[t]
 \centering
 \includegraphics[width=\columnwidth]{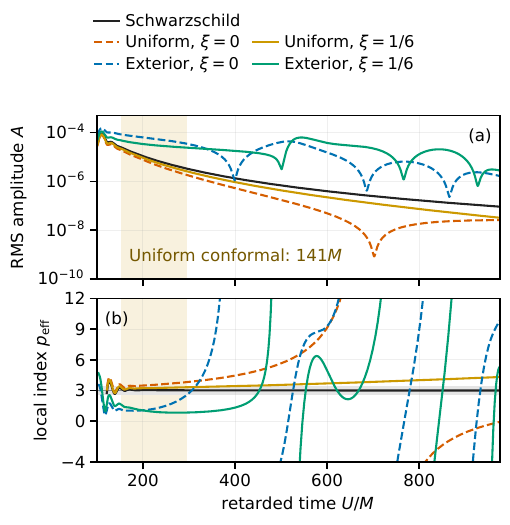}
 \caption{Matched dipole tails at $L/M=640$, compared at $\hc$ with
 Schwarzschild at $\scri^+$.  (a) Centered $10M$ RMS envelopes.
 (b) Local power indices from $40M$ fits; the gray band marks
 $p=3\pm10\%$ and the gold shading the uniform-conformal Price-compatible interval.}
 \label{fig:matched-tail}
\end{figure}

At $L/M=640$, the exterior-supported construction fails to recover the
Schwarzschild tail decay rate for either coupling
(Fig.~\ref{fig:matched-tail}).  Both evolutions develop slowly decaying
components and secondary lobes that persist under refinement, and neither
meets the $40M$ criterion.  This behavior is consistent with the sensitivity
of tails to scattering in the outer geometry, which remains modified by the
regulator.  Appendix~\ref{app:matched-tail-protocol} gives the intervals and
convergence evidence.

\section{Conclusions}
\label{sec:conclusions}

We tested Misner's artificial cosmology as a regulator at the conformal boundary using scalar radiation from a compact source in a black-hole spacetime.  The regulator introduces a small positive cosmological constant and a smooth foliation that reaches the cosmological horizon.  The resulting SdS spacetime is asymptotically de Sitter, with a conformal boundary at finite radius. An important aspect of the construction is that the foliation must resolve the conformal boundary of the cosmological spacetime for it to have a numerically useful limit to the asymptotically flat case (see Fig.~\ref{fig:bridge_foliations} and \cite{zenginoglu_bridge_2025, ZhouPanossoMacedo:2025}). Then, the limiting slices meet distinct cuts of future null
infinity, and their conformal spatial metric remains smooth and
nondegenerate.  In this paper, we have used a minimal-gauge foliation that recovers the Schwarzschild slices in the limit of vanishing cosmological constant
\cite{AnsorgMacedo:2016,Macedo:2020,PanossoMacedo:2023qzp,ZhouPanossoMacedo:2025}.

Beyond testing Misner's numerical regulator on wave propagation, these calculations provide a model for understanding the effects of a small positive cosmological constant on outgoing waveforms.  The single-mode and localized-source comparisons show how Schwarzschild recovery improves with increasing $L$ and provide benchmarks for choosing a cosmological scale.  At sufficiently large $L$, we demonstrated that uniform SdS provides accurate ringdown waveforms and even gives the polynomial Price-law decay rate in intermediate intervals. The coupling comparison shows that tail recovery depends on
the asymptotic field equation.
This construction can also be used to numerically test the asymptotically flat approximation for isolated systems in a Universe with a small positive cosmological constant
\cite{Ashtekar:2014zfa,Ashtekar:2015lxa}.

We extend Misner's artificial cosmology by restricting the cosmological modification to an exterior layer, based on Dymnikova's conserved anisotropic source \cite{Dymnikova:2000algebraic,Dymnikova:2002mass}. This approach preserves the strong-field geometry and improves ringdown agreement relative to uniform SdS at fixed $L$.
However, it does not improve tail recovery.
The transition layer changes the scattering potential, and tails are sensitive to the outer geometry and source history
\cite{DeAmicisEtAl:2024tails,RosatoEtAl:2026GreenFunction}.
Extending this construction to nonlinear Einstein evolution is more
difficult.  Source conservation must hold as the metric evolves, requiring
a covariant evolution law beyond the static spherical ansatz.  The stresses in the transition layer also violate the null energy condition.

A natural next step is to compute gravitational waveforms from particle
sources inspiraling into the central black hole.  This would test
cosmological effects on long, astrophysically relevant signals.  A separate
nonlinear Einstein study is needed to determine whether Misner's artificial cosmology
is a practical proposal for the treatment of the asymptotic boundary.

\begin{acknowledgments}
A.Z. thanks Ted Jacobson, Rodrigo Panosso Macedo, and Alex
Va\~n\'o-Vi\~nuales for discussions.  This work was inspired by Charles
Misner's investigations of artificial cosmology and by the memorial
symposium held in his honor at the University of Maryland in 2023.  This material is
supported by NSF Grant No.~2309084. A.Z. used OpenAI Codex and G.A.K. used Anthropic Claude to assist with code
development, numerical analysis, and manuscript review.
\end{acknowledgments}

\appendix

\section{Numerical verification}
\label{app:convergence}

This appendix details the numerical evidence supporting the waveform
comparisons.  Tail-specific criteria and refinement results are 
in Appendix~\ref{app:latetime}.

Independent implementations give consistent results for the sourced response
and exterior ringdown.  Finite-difference and spectral sourced waveforms on
Schwarzschild and uniform SdS at $L/M=80$ agree to $5.58\times10^{-4}$
in the relative sphere-integrated norm.  At $L/M=320,640$, the exterior
ringdown waveforms from the spectral-element solver and an independent
first-order calculation agree to $1.12\times10^{-5}$ in the
Schwarzschild-normalized $L^2$ norm on $15\leq U/M\leq45$.
These differences are small compared with the reported finite-$L$ effects;
the exterior-supported improvement persists under spatial and temporal
refinement.

\subsection{Fixed-mode and localized-source convergence}

The minimally coupled uniform fixed-mode sequence uses
$(N,\Delta\tau/M)=(384,0.005)$, $(512,0.00375)$, and $(768,0.0025)$,
evolved to $\tau=200M$ with boundary output every $0.03M$.  The
localized-source sequence uses
$(N_r,\Delta\tau/M,\ell_{\max})=(1024,0.001,42)$,
$(1536,1/1500,46)$, and $(2048,0.0005,50)$, evolved to $\tau=60M$
with output at every timestep.  Each finite-$L$ run has a separately
evolved Schwarzschild control at the same resolution.  Radial resolution,
timestep, and, for the localized source, angular truncation are refined
together.

We test convergence of waveform differences directly.  For a residual
$R_q=X_q-Y_q$ at refinement level $q$, the medium-to-fine change is
\begin{equation}
 \delta_{mf}=
 \frac{\|R_{\rm medium}-R_{\rm fine}\|}
 {\|W_{{\rm Schw},{\rm fine}}\|},
 \label{eq:paired_refinement_change}
\end{equation}
using the fixed-mode or Parseval norm as appropriate. Richardson estimates use an effective refinement
parameter $h_q=N_q^{-1}$ and an order $p$ inferred from the two successive
changes with their actual resolution ratios.  This order describes the
joint refinement.  The estimated remaining fine-grid error is
$\delta_{mf}/[(N_f/N_m)^p-1]$; the larger of this estimate and the observed
change sets the numerical scale used for the direct accuracy thresholds.
These tests quantify refinement at fixed coefficient and matrix cutoffs.
An independent cutoff check for the large-$L$ tail is given in
Appendix~\ref{app:large-L-tail}.

For the extrapolants, refinement changes in all three input waveforms
propagate into the residual comparisons.  Table~\ref{tab:extrapolants}
summarizes the largest values over the three cumulative intervals.  The
observed changes reach $0.409\%$, exceeding the central residuals near
$0.03\%$ and supporting the conservative $1\%$ agreement stated in
Sec.~\ref{sec:direct_recovery}.

\begin{table}[t]
 \caption{Minimally coupled extrapolation checks.  Each entry is the
 maximum over the cumulative intervals ending at $U/M=40,80,160$.
 Fine-grid residuals, observed medium-to-fine residual changes, and
 propagated Richardson estimates are reported separately, in percent.}
 \label{tab:extrapolants}
 \begin{ruledtabular}
 \begin{tabular}{lrrr}
 comparison & residual & change & estimate\\
 \colrule
 $W_\infty^{(80)}-W_{\rm Schw}$ & 0.0222 & 0.2666 & 0.0777\\
 $W_\infty^{(160)}-W_{\rm Schw}$ & 0.0265 & 0.1430 & 0.0413\\
 $W_\infty^{(80)}-W_\infty^{(160)}$ & 0.0313 & 0.4092 & 0.1189\\
 \end{tabular}
 \end{ruledtabular}
\end{table}

The cumulative norms are dominated by the early signal.  On the disjoint
interval $40\leq U/M\leq80$, extrapolant residuals of $0.096$--$0.211\%$
have refinement changes of $1.02$--$3.14\%$.  Refinement also exceeds the
residuals on $80\leq U/M\leq160$.

Localized-source refinement changes remain at the $10^{-4}\%$ level on
the common simulated interval.  For the Schwarzschild control, the
sphere- and time-integrated relative differences from the fine solution are
$3.44\times10^{-7}$ and $7.66\times10^{-8}$ at coarse and medium resolution.
Truncating the fine solution at $\ell_{\max}=42$ and $46$ gives omitted-mode
norms of $2.29\times10^{-9}$ and $1.26\times10^{-10}$, respectively.
All four values use the fine solution's Parseval norm for normalization.

\subsection{Exterior ringdown}
\label{app:ringdown-verification}

The minimally coupled quadrupole results in
Table~\ref{tab:exterior-qnm-improvement} use spectral-element calculations through
$U=80M$ with degrees $24,32,40$ at $\Delta\tau=0.05M$, and a degree-$32$
check at $0.025M$.  All three geometries use the displacement data
\eqref{eq:displacement_data} and the same geometric clock.  Reported errors
use degree $40$ and a common fine Schwarzschild reference.

On $15\leq U/M\leq45$, the largest successive spatial changes of the
paired residuals are $3.05\times10^{-9}$ and $6.73\times10^{-11}$;
timestep halving gives $5.79\times10^{-8}$.  The largest individual waveform
change under timestep halving is $1.96\times10^{-6}$.  All are $L^2$
fractions normalized by the fine Schwarzschild waveform.  Replacing the
common reference in both errors by the half-step Schwarzschild solution
changes $\Delta E=E_{\rm SdS}-E_\chi$ by at most $2.44\times10^{-8}$.
These changes are small compared with the measured improvements
$\Delta E=0.00408$ and $0.00134$ at $L/M=320$ and $640$.

\section{Tail decay and numerical accuracy}
\label{app:latetime}

We describe how the tail decay rates are measured and checked for
numerical accuracy.

\subsection{Schwarzschild tail at large \texorpdfstring{$L$}{L}}
\label{app:large-L-tail}

At $L/M=3072$, we compare minimally coupled SdS at $\hc$ with Schwarzschild
at $\scri^+$, using the same compact $\ell=1$ velocity data and geometric
clock.  We use $N=1536,2048,3072$ at $\Delta\tau=0.0025M$ and halve the
timestep at $N=2048$.

We measure the local power index by fitting $\ln A$ against $\ln U$ over
$40M$, using the $10M$ RMS envelope \eqref{eq:rms-envelope}.  Both indices
must stay within $5\%$ of $p=3$ and satisfy
$|p_{\rm SdS}-p_{\rm Schw}|\leq0.15$ for a continuous $150M$ interval
containing $U=300M$.  The two finest resolutions must also agree in envelope
amplitude to within $1\%$.  The reported interval ends at departure from
this criterion: $232\lesssim U/M\lesssim382$.
Table~\ref{tab:large-L-tail-convergence} shows the changes under refinement.

\begin{table}[t]
 \caption{Largest relative envelope changes on the $L/M=3072$ Price-law
 interval, $232\lesssim U/M\lesssim382$.  Values are fractions for SdS at
 $\hc$ and Schwarzschild at $\scri^+$.  The timestep check uses $N=2048$.}
 \label{tab:large-L-tail-convergence}
 \begin{ruledtabular}
 \begin{tabular}{lcc}
 refinement & SdS & Schwarzschild\\
 \colrule
 $N:1536\to2048$ & $1.27\times10^{-4}$ & $3.58\times10^{-4}$\\
 $N:2048\to3072$ & $4.77\times10^{-5}$ & $2.20\times10^{-5}$\\
 $\Delta\tau/M:0.0025\to0.00125$ & $3.29\times10^{-8}$ & $3.36\times10^{-8}$\\
 \end{tabular}
 \end{ruledtabular}
\end{table}

Departure occurs near $U=382M$ and varies by about $2M$ with spatial
resolution.  Halving the timestep leaves it unchanged.  This spread is
measured with the decay-rate estimator held fixed.

A further check at $N=2048$ and $\Delta\tau=0.0025M$ lowers the truncation
thresholds for coefficients and matrices from $(10^{-6},10^{-12})$ to
$(10^{-10},0)$.  It extends to $U=450M$ and also uses a different computing
platform.  The largest relative envelope change is $2.5\times10^{-4}$,
exceeding the finest spatial-refinement change; the largest absolute index
change is $3.5\times10^{-4}$.  Departure moves about $0.8M$ earlier, leaving
the Price-law conclusion unchanged.

The main runs extend to $U/M\simeq1984$, or $\kappa_cU=0.645$.  To identify
cosmological decay, we require the envelope to exceed ten times the
numerical error estimated from refinement. The
intermediate decay rate agrees with Schwarzschild to within $5\%$.

\subsection{Uniform and exterior-supported tails}
\label{app:matched-tail-protocol}

The $L/M=640$ comparison in Fig.~\ref{fig:matched-tail} uses the
spectral-element method, with degrees
$24,32,40$ at $\Delta\tau=0.1M$ and a degree-$40$ run at $0.05M$, through
$U=1000M$.  Schwarzschild and uniform SdS use nine elements; the exterior
cases use sixteen.  The figure uses degree $40$ and the smaller timestep,
with the same Schwarzschild reference for both couplings.

Using the same envelope and fit as above, we require $p_{\rm eff}$ to
remain within $10\%$ of $3$ for at least $40M$.  We retain an index only
where the envelope exceeds ten times the numerical error estimated from
refinement and the index changes by less than $0.1$ under both degree-$32$
to $40$ refinement and timestep halving.  We omit times within half a fit
window of a waveform zero.  Envelope widths $5M$, $10M$, and $20M$, combined
with fit widths $30M$, $40M$, and $60M$, leave the outcome of this test
unchanged.

Only uniform conformal SdS meets the $40M$ requirement among the finite-$L$
cases, on $153\lesssim U/M\lesssim295$
(Table~\ref{tab:matched-tail-intervals}).  Both exterior cases fail the test.
Spatial differences decrease from the degree-$24$--$32$ comparison to the
degree-$32$--$40$ comparison in every case.  The agreement concerns the
decay rate alone.

\begin{table}[t]
 \caption{Longest intervals with $p_{\rm eff}=3\pm10\%$ that pass the
 numerical checks, for $\ell=1$ and $L/M=640$.  The required duration is
 $40M$; the Schwarzschild interval ends with the usable data.  The waveform
 changes $\delta_s$ and $\delta_t$ are relative $L^2$ norms.}
 \label{tab:matched-tail-intervals}
 \setlength{\tabcolsep}{4pt}
 \begin{ruledtabular}
 \begin{tabular}{lrrr}
 case & $\Delta U/M$ & $\delta_s$ & $\delta_t$\\
 \colrule
 Schwarzschild & $\geq834.6$ & $4.59\times10^{-11}$ & $1.58\times10^{-7}$\\
 uniform, $\xi=0$ & $9.6$ & $1.82\times10^{-10}$ & $2.18\times10^{-7}$\\
 uniform, $\xi=1/6$ & $141.5$ & $1.82\times10^{-10}$ & $1.83\times10^{-7}$\\
 exterior, $\xi=0$ & $13.3$ & $2.43\times10^{-7}$ & $3.06\times10^{-8}$\\
 exterior, $\xi=1/6$ & $8.5$ & $3.52\times10^{-6}$ & $1.16\times10^{-7}$\\
 \end{tabular}
 \end{ruledtabular}
\end{table}

Here $\delta_s$ compares degrees $32$ and $40$; $\delta_t$ measures the
change on halving the timestep.  Each is divided by the finer waveform norm
and maximized over $150$--$300M$, $300$--$500M$, $500$--$750M$, and
$750$--$950M$ for the same geometry and coupling.

For cosmological decay, we measure $\gamma_{\rm eff}=-d\ln A/dU$.
At large $L$, uniform SdS has the expected rate \cite{brady_radiative_1999}
\begin{equation}
 \frac{\gamma_{\ell,\xi}}{\kappa_c}
 =\ell+\frac32-\frac12\sqrt{9-48\xi}
 +\mathcal O(M/L),
 \label{eq:brady_exponent}
\end{equation}
giving dipole rates $\gamma_{1,0}/\kappa_c\simeq1$ and
$\gamma_{1,1/6}/\kappa_c\simeq2$.  Neither uniform waveform stays within
$10\%$ of its expected rate for a full exponential-fit interval
$\Delta(\kappa_cU)=0.25$ before the runs end at $\kappa_cU\simeq1.56$.
Cosmological decay is also unresolved for the exterior profiles.

\subsection{Cosmological decay at finite radius}
\label{app:finite-radius-crossover}

At $r=8M$, we compare minimally coupled $\ell=1$ signals at $L/M=80,160$
with Schwarzschild. Departure ends the last sustained
agreement with Schwarzschild before cosmological entry; entry begins the
first sustained agreement with the expected SdS rate.  We vary smoothing,
envelope width, required duration, and tolerance
over 54 settings, requiring both times to be found in at least half of them.
Quoted medians and ranges use the settings that identify both times.

For $L/M=80,160$, the entry ranges are $\kappa_cU\in[2.56,3.07]$ and
$[2.68,3.19]$; the departure ranges are $[1.15,1.76]$ and $[0.71,0.96]$.
These ranges accompany the medians quoted in Sec.~\ref{sec:sds_physics}.
Both departures precede the settled Schwarzschild Price tail, which begins
near $U=220M$ under the $5\%$ criterion.  The observed transition thus
connects a Schwarzschild-like transient to cosmological decay.

\section{Exterior profile and boundary limits}
\label{app:exterior-implementation}

The exterior profile keeps the transition layer and outer exact-SdS layer at nonvanishing compactified widths.  We specify this profile and derive its geometric,
boundary-coefficient, and clock limits. 

\subsection{Transition profile and large-\texorpdfstring{$L$}{L} limits}

Using \eqref{eq:exterior_compactification}, define the Chebyshev angle
$\vartheta(r)=\arccos[2\rho_\chi(r)-1]$ on $2M\leq r\leq r_c$.
Chebyshev collocation points are uniformly spaced in $\vartheta$.
For $\beta=9/10$, set
\begin{equation}
\begin{aligned}
 \widehat R_1(L)&=\beta r_c,\\
 \widehat\rho_{\chi,1}(L)&=
 \frac{1-2M/\widehat R_1}{D_L},\\
 \widehat\vartheta_1(L)&=
 \arccos\!\left(2\widehat\rho_{\chi,1}-1\right).
\end{aligned}
 \label{eq:horizon_scaled_transition}
\end{equation}
The transition edges obey
\begin{equation}
\begin{aligned}
 \frac{L_\star}{M}&=160,\\
 \vartheta_\star&=\widehat\vartheta_1(L_\star)
 \simeq0.07526,\\
 \vartheta_1(L)&=
 \max\{\widehat\vartheta_1(L),\vartheta_\star\},\\
 \vartheta_0(L)&=2\vartheta_1(L),\\
 \rho_{\chi,i}(L)&=\frac{1+\cos\vartheta_i(L)}{2},\\
 R_i(L)&=\frac{2M}{1-D_L\rho_{\chi,i}(L)}
 \quad(i=0,1).
\end{aligned}
 \label{eq:transition_radii}
\end{equation}
Thus $2M<R_0<R_1<r_c$, and the transition and outer layers have equal angular widths,
$\vartheta_0-\vartheta_1=\vartheta_1$.  For $L\leq L_\star$, $R_1=\beta r_c$;
the $L/M=80,160$ members retain the horizon-scaled profile.  For
$L>L_\star$, both widths equal $\vartheta_\star$, with fixed edge coordinates
$\rho_{\chi,0}^\star=(1+\cos2\vartheta_\star)/2$ and
$\rho_{\chi,1}^\star=(1+\cos\vartheta_\star)/2$.  The compactified widths are
\begin{align}
 \Delta\rho_{\rm tr}
 &=\rho_{\chi,1}^\star-\rho_{\chi,0}^\star
 \simeq0.004239,\notag\\
 \Delta\rho_{\rm layer}
 &=1-\rho_{\chi,1}^\star\simeq0.001416.
 \label{eq:transition_compact_widths}
\end{align}
The smooth switch \eqref{eq:exterior_switch} uses
\begin{equation}
 \zeta_L(r)=\frac{\vartheta_0-\vartheta(r)}
 {\vartheta_0-\vartheta_1}.
 \label{eq:production_exterior_switch}
\end{equation}
At the two tested lengths above $L_\star$, the radii are
\begin{equation}
\begin{array}{c|cc}
 L/M & R_0/M & R_1/M\\ \hline
 320 &168.2&260.5\\
 640 &228.1&440.4
\end{array}
 \label{eq:fixed_angle_transition_radii}
\end{equation}
Their finite limits are
\begin{equation}
\begin{aligned}
 \frac{R_{0,\infty}}{M}
 &=\frac{2}{1-\rho_{\chi,0}^\star}\simeq353.7,\\
 \frac{R_{1,\infty}}{M}
 &=\frac{2}{1-\rho_{\chi,1}^\star}\simeq1412.9,\\
 \frac{R_0}{r_c}&\longrightarrow0,
 &\frac{R_1}{r_c}&\longrightarrow0.
\end{aligned}
 \label{eq:transition_radii_limit}
\end{equation}
The transition therefore approaches a finite, distant interval, while the
exact-SdS layer extends from its outer edge to $r_c$.  Since $R_0>6M$
throughout the numerical sequence, the strong-field region remains exactly
Schwarzschild.  Equation~\eqref{eq:transition_mass} gives
\begin{equation}
 \frac{\Delta m_{\rm tr}}{M}
 \sim\frac12\left(\frac{R_{1,\infty}}{M}\right)^3
 \left(\frac{M}{L}\right)^2\longrightarrow0.
 \label{eq:transition_mass_fixed_angle_family}
\end{equation}

For an areal width $\Delta=R_1-R_0$, the derivative scales are
\begin{equation}
 \lambda'=\mathcal O\!\left(\frac{\Lambda}{\Delta}\right),
 \qquad
 \lambda''=\mathcal O\!\left(\frac{\Lambda}{\Delta^2}\right).
 \label{eq:transition_derivative_scales}
\end{equation}
Narrow layers can generate large curvature and sharp scattering features.
Here $\Delta\to R_{1,\infty}-R_{0,\infty}\simeq1059.2M$, so the limiting
profile has $\chi_L'=\mathcal O(M^{-1})$ and
$\chi_L''=\mathcal O(M^{-2})$.  For $\lambda=3\chi_L/L^2$ at fixed $M$,
\begin{equation}
 \lambda'=\mathcal O(L^{-2}M^{-1}),
 \qquad
 \lambda''=\mathcal O(L^{-2}M^{-2}).
 \label{eq:transition_width_floored_derivative_scales}
\end{equation}
The source eigenvalues and Ricci curvature in the transition are therefore
$\mathcal O(L^{-2})$, although the lapse change near $r_c$ remains order
unity.

For conformal coupling, the potential \eqref{eq:scalar_potential} becomes
\begin{equation}
 V_{\ell,1/6}^\chi=f_\chi\left[
 \frac{\ell(\ell+1)}{r^2}+\frac{2M}{r^3}
 +\frac{r\lambda'}{9}+\frac{r^2\lambda''}{18}\right].
 \label{eq:exterior_conformal_scalar_potential}
\end{equation}
It equals the Schwarzschild potential below $R_0$ and the corresponding
SdS potential above $R_1$.  

\subsection{Boundary coefficients and clock asymptotics}

The coefficients follow from Eqs.~\eqref{eq:coefficients} and
\eqref{eq:potential_coefficient} with $G_\chi=2M/(D_Lr^2)$; removable
horizon limits are assigned analytically using \eqref{eq:A_endpoint}.
The bridge crosses both horizons regularly:
\begin{align}
 \lim_{r\to2M^+}\frac{1-B_\chi^2}{f_\chi}&=8,\notag\\
 \lim_{r\to r_c^-}\frac{1-B_\chi^2}{f_\chi}
 &=\frac{16M^2}{r_c(r_c-3M)}>0.
 \label{eq:exterior_bridge_regular_limits}
\end{align}
In the exact-SdS layer, $D_L=r_c^2/L^2=1-2M/r_c$, giving
\begin{align}
 A_\chi(1)&=\frac{r_c-3M}{8M(r_c-2M)},&
 c_{+,\chi}(1)&=2A_\chi(1),\notag\\
 P^\chi_{\ell,\xi}(1)&=\frac{D_L}{2M}
 \left[\ell(\ell+1)-2+\frac{6M}{r_c}\right.\notag\\
 &\left.\hspace{7em}+12\xi D_L\right].
 \label{eq:exterior_outer_endpoint_coefficients}
\end{align}
Their flat limits are
\begin{align}
 A_\chi(1)&\longrightarrow\frac{1}{8M},&
 c_{+,\chi}(1)&\longrightarrow\frac{1}{4M},\notag\\
 P^\chi_{\ell,0}(1)&\longrightarrow
 \frac{\ell(\ell+1)-2}{2M},&
 P^\chi_{\ell,1/6}(1)&\longrightarrow
 \frac{\ell(\ell+1)}{2M}.
 \label{eq:exterior_outer_endpoint_flat_limit}
\end{align}
For minimal coupling, convergence of the lower-order potential near the
outer boundary remains nonuniform.  With $\delta=2M/r_c$, the potential in
the outer layer contains
\begin{equation}
 -\frac{D_Lr^2}{ML^2}
 =-\frac{D_L^2}{M}
 \left[\frac{\delta}{\delta+D_L(1-\rho_\chi)}\right]^2.
 \label{eq:exterior_cap_boundary_layer}
\end{equation}
Its boundary amplitude is order $M^{-1}$, while its compactified width
shrinks as $M/L$.  Fixed geometric widths thus do not ensure accuracy at
arbitrarily large $L$ and fixed resolution.  Conformal coupling cancels
this term; the transition terms still require resolution checks.

The clock integrand has the finite horizon limit
\begin{equation}
 \lim_{r\to r_c}\frac{1+B_\chi}{f_\chi}
 =\frac{B_\chi'(r_c)}{f_\chi'(r_c)}
 =\frac{8M^2}{r_c(r_c-3M)}.
 \label{eq:exterior_clock_endpoint}
\end{equation}
For its flat limit, choose a fixed $R>R_{1,\infty}$.  At sufficiently large
$L$, $[R,r_c]$ is exactly SdS, while on $[r_0,R]$ the integrand in
\eqref{eq:exterior_clock} differs from its Schwarzschild value by
$\mathcal O(L^{-2})$.  In the outer layer, a cancellation-free form is
\begin{equation}
 \frac{1+B_\chi}{f_\chi}
 =\frac{8M^2(r_c+r)}{r^2r_c^2
 \left[(r_c+r)/L^2-2M/(rr_c)\right]}.
 \label{eq:exterior_clock_cap}
\end{equation}
Its integrated difference from the Schwarzschild integrand is
$\mathcal O[M^3\log(r_c/R)/r_c^2]$, while the omitted Schwarzschild tail is
$8M^2/r_c+\mathcal O(M^3/r_c^2)$.  Together these give
$q_\chi=q_0-8M^2/r_c+o(M^2/r_c)$, with the same normalization
$q_0=4M\ln[r_0/(r_0-2M)]$ as in
\eqref{eq:exterior_clock_flat_limit}.

\bibliographystyle{apsrev4-2}
\makeatletter
\immediate\write\@auxout{\string\citation{apsrev42Control}}
\makeatother
\bibliography{SdS_refs}

\end{document}